\documentclass{emulateapj}

\def\gsim{\;\rlap{\lower 2.5pt
 \hbox{$\sim$}}\raise 1.5pt\hbox{$>$}\;}
\def\lsim{\;\rlap{\lower 2.5pt
   \hbox{$\sim$}}\raise 1.5pt\hbox{$<$}\;}

\begin{document}

\title{Breaking Cosmological Degeneracies in Galaxy Cluster Surveys\\with a Physical Model of Cluster Structure}
\author{Joshua D. Younger\altaffilmark{1}, Zolt\'{a}n Haiman\altaffilmark{2}, Greg L. Bryan\altaffilmark{2}, and Sheng Wang\altaffilmark{3,4}}
\vspace{\baselineskip}
\affil{$^1$Columbia Astrophysics Lab, Columbia University, Pupin Physics Laboratories, New York, NY 10027}
\affil{$^2$Department of Astronomy \& Astrophysics, Columbia University, Pupin Physics Laboratories, New York, NY 10027}
\affil{$^3$Department of Physics, Columbia University, Pupin Physics Laboratories, New York, NY 10027}
\affil{$^4$Brookhaven National Laboratory, Upton, NY11973--5000}

\begin{abstract}
It has been shown that in an idealized galaxy cluster
survey, containing $\gsim 10,000$ clusters, statistical errors on dark
energy and other cosmological parameters will be at the percent level.
Furthermore, through ``self--calibration'', parameters
describing the mass--observable relation and cosmology can be
simultaneously determined, though at a loss in accuracy by about an
order of magnitude. Here we examine an alternative approach to
self--calibration, in which a parametrized ab--initio physical model
is used to compute theoretical mass--observable relations from the cluster structure.  
As an example, we use a modified--entropy
(``preheating'') model of the intracluster medium, with the history and magnitude
of entropy injection as unknown input parameters. Using a Fisher
matrix approach, we evaluate the expected simultaneous statistical
errors on cosmological and cluster model parameters.  We find that compared to a
phenomenological parametrization of the mass--observable relation, our
physical model yields significantly tighter constraints in both
surveys, and offers substantially improved synergy when the two
surveys are combined.  In a mock X--ray survey, we find statistical
errors on the dark energy equation of state are a factor of two tighter than the phenomenological model
,  with $\Delta w_0 \sim 0.08$ and its evolution, $\Delta w_a \equiv -\Delta dw/da \sim 0.23$, with
corresponding errors of $\Delta w_0 \sim 0.06$ and $\Delta w_a \sim
0.17$ from a mock Sunyaev-Zel'dovich (SZ) survey, both with $N_{cl} \sim 2.2\times 10^4$ clusters, while
simultaneously constraining cluster model parameters to $\lsim 10\%$.
When the two surveys are combined, the constraints tighten to $\Delta
w_0 \sim 0.03$ and $\Delta w_a \sim 0.1$; a 40\%
improvement over adding the individual experiment errors in
quadrature, and a factor of 2 improvement over the phenomenological model.  
This suggests that parametrized physical models
of cluster structure will be useful when extracting cosmological
constraints from SZ and X--ray cluster surveys.
\end{abstract}

\section{Introduction}

It has been known for about two decades that clusters of galaxies
provide a uniquely powerful probe of fundamental cosmological
parameters.  As the most massive virialized objects in the universe,
they form at the highest peaks in the primordial density field, and as
a result, their abundance and spatial distribution are very sensitive
to the underlying cosmology
\citep[e.g.,][]{bahcall1988,peebles1993,carlberg1997,rosati2002,henry2004}.
Strong constraints have already been derived on the present day matter
density $\Omega_m$, and the normalization of the dark matter power
spectrum $\sigma_8$
\citep{peebles1989,henry1991,bahcall1992,viana1996,bahcallbode2003,bahcall2003},
using only dozens of clusters.  In large future surveys, with tens of
thousands of clusters, high--precision constraints are expected to be
available on dark energy parameters \citep{haiman2001}, including
constraints on the evolution of the equation of state $w_a\equiv
dw/da$ \citep{weller2002,weller2003,wang2004}.

Plans are are currently being drafted for the first very large scale
clusters surveys.  These projects, such as the 4,000 degree$^2$ South
Pole Telescope that will find clusters through the SZ effect
(hereafter SPT; Ruhl et al. 2003), a proposed larger X-ray survey
\citep{haiman2005}, or catalogs of clusters identified in wide
weak--lensing surveys, such as LSST \citep{lsst}, promise to produce
catalogs of many thousands of clusters, containing a rich resource of
cosmological information.  With such an increase in the data looming
on the horizon, it is becoming increasingly important to understand
the inherent systematics, and establish realistic expectations for
potential cosmological constraints.

Several recent works, motivated by the above considerations, have
focused on various aspects of extracting cosmological parameters from
high--yield future surveys, such as the statistical constraints
available on curvature $\Omega_k$~\citep{holder2001}, assessing the
impact of sample variance~\citep{hukrav2003} and other
uncertainties~\citep{levine2002} on parameter estimates, and
controlling such uncertainties by utilizing information from the shape
of the cluster mass function $dN/dM$~\citep{hu2003}.  Recent studies
have also elucidated the additional cosmological information available
from the spatial distribution of galaxy clusters through a measurement
of their three--dimensional power spectrum \citep{huhaiman2003},
utilizing both the intrinsic shape of the transfer
function~\citep{refregier2002} and baryon
features~\citep{blake2003,seo2003,linder2003}.

Most importantly, recent work has focused on the systematic errors on
cosmological parameters arising from the inherent uncertainties on the
mass--observable relation.  It has been shown that large forthcoming
cluster surveys can be ``self--calibrating'', in the sense that when
the abundance and clustering properties of clusters are considered in
tandem, tight constraints can be derived on cosmology even if the
mass--observable relation has to be determined simultaneously from the
same data \citep{majumdar2004, wang2004, lima2005}.  In these studies,
the mass--observable relation was assumed to be either unknown
apriori, or to have a simple parametric form (such as a power--law).

While these assumptions have been demonstrated to yield interesting
constraints, one may argue that this approach is overly conservative.
In particular, it is reasonable to expect that the structure of
clusters, such as their mean density and temperature profiles, can be
at least approximately computed from ab--initio models.  The simplest, spherically symmetric, self--similar models contradict
observations, such as the relation between X--ray flux ($L_x$) and
temperature ($T$), or the resolved profiles of low--mass clusters \citep{voit02}.
However, motivated by the observations of \citet{ponman1999},
modified--entropy models of the intra--cluster medium (ICM), while
still not correct in detail \citep{pratt2005,pratt2006,younger2006}, have been
successful at predicting the mass--observable scalings in temperature,
luminosity, and SZ effect, from first principles
\citep{voit02,mccarthy2003a,mccarthy2003b}.  It is reasonable to
expect that our understanding of the physics determining the structure
of clusters will be improved in the future by advances in both
observations and numerical simulations.

In this paper, we study the utility of a partial understanding of the
ICM in improving cosmological constraints.  We will show that using
a--priori cluster structure models, it is possible to significantly
improve cosmological constraints, relative to direct, phenomenological
self--calibration of the mass--observable relation -- even if the physical model has free parameters.
Furthermore, the combination of X-ray and SZ datasets is particularly promising \citep{verde2002}, and can more
effectively break degeneracies, since the underlying models directly
relate the SZ decrement to the X--ray flux. The end result is
significantly improved errors on the cosmology, using the same
observational data set, in addition to a well constrained model of the
ICM.

The rest of this paper is organized as follows.  
In \S\ref{sec:models}, we
describe our models of cluster structure and evolution in 
the presence of a preheating of the ICM gas.
In \S\ref{sec:fiducial}, we
discuss our fiducial model, and demonstrate that it satisfies a number
of existing observations.
In \S\ref{sec:selfcal}, we
contrast our physical models with a ``traditional'' conservative
approach to self--calibration, in which the mass--observable relation
is parametrized directly as a power--law.
In \S\ref{sec:surveys}, we
discuss and motivate the parameters of our mock cluster surveys, and, for
comparison, discuss future 
measurements of cosmic microwave background (CMB) anisotropies.
In \S\ref{sec:fisher}, we
briefly summarize the Fisher matrix approach, as applied to cluster surveys and CMB anisotropies.
In \S\ref{sec:results}, we
present and critically discuss our results, by comparing the constraints obtained under
self--calibration and under the assumption of physical cluster models.
Finally, in \S\ref{sec:conclude}, we summarize our main conclusions and the implications of this work.

%\newpage

\section{Cluster Structure and Evolution Models}
\label{sec:models}

In this section, for completeness and specificity, we describe our models for the structure -- density
and temperature profiles -- of galaxy clusters, in the presence or
absence of excess entropy.  Our treatment assumes that the gas is in
hydrostatic equilibrium in a dark matter (DM) potential, and it closely
follows previous studies on cluster profiles and preheating.  

\subsection{Motivation}

Modified entropy models are physically motivated by the presence of a
universal entropy floor.  \citet{ponman1999} find that the entropy
distribution of ICM gas, as inferred from ROSAT observations of 25
galaxy clusters, is inconsistent with simple gravitational collapse,
and suggest instead the presence of an entropy floor of $K_0 \sim 100
h^{-1/3}$ keV cm$^2$.  This excess entropy is itself potentially a
relic of winds and outflows from galaxies that formed in the
protocluster, and heated the ICM at moderate redshift $z\sim 2$, prior
to cluster collapse.  They then argue that this results in the
observed departure of real galaxy clusters from the expected
self-similar scaling.  However, it is plausible that only a fraction
of this preheating entropy is injected at high redshift.  Some of the
heat input may come from a combination of active galactic nuclei (AGN)
and supernovae (SN), over an extended period, rather than
winds in a narrow redshift range.  For this reason, it is useful to
treat $K_0$ as a function of redshift, which can be parametrized, for
example, by a power-law relation $K_0(z) = K_0(z=0)(1+z)^{\alpha_K}$.

We assume that clusters at a given redshift $z$ are spherically
symmetric and in hydrostatic equilibrium, with an ideal gas ICM
consisting of a fully ionized H--He plasma with the cosmic helium mass
fraction $Y_{\rm He}=0.25$, and a mean molecular weight $\mu=0.59$.
The unmodified entropy distribution is constructed by assuming that
the gas traces the dark matter (DM), and that the total gas density is
given by a fraction $f_g$ of the cosmological baryon density within
the virialized cluster that remains in the ICM.  The DM and gas
profile is assumed to follow the \citet{nfw97} (NFW) form, as
motivated by N-Body simulations, with a fixed concentration parameter
$c=5$ \citep{eke2001}.  The assumption of hydrostatic
equilibrium, together with physically motivated boundary conditions,
allows us to calculate the pressure and temperature profiles, which,
combined with the gas density, give the entropy distribution as a
function of radius.  This entropy distribution is then modified by
adding a constant $K_0$ at all radii, and the hydrostatic equilibrium
equations are re--integrated under the assumption that mass shells do
not cross (i.e. $K(M_g)$ is conserved where $M_g$ is the total gas
mass at that radius).  The resulting modified density and temperature
profiles have flat cores, and they can be used to predict cluster
observables, such as the X--ray flux or SZ decrement.

\subsection{Unmodified Equilibrium Structure}

The ICM is assumed to trace the DM, which is well-described
by an NFW profile of the form
\begin{equation}
\rho_{DM}(r) = \frac{\delta_c \rho_c}{(r/R_s)(1+ r/R_s)^2}
\end{equation}
where $R_s$ is a scale radius defined in terms of the concentration
parameters $c$ and the virial radius, as in \citet{bullock2001},
by
\begin{equation}
c R_s = R_{v} =
\left(\frac{3M}{4\pi\Delta_v\rho_c}\right)^{1/3}
\end{equation}
and
\begin{equation}
\delta_c =
\frac{\Delta_v}{3}\frac{c^3}{\ln{(1+c)}-\frac{c}{1+c}}.
\end{equation}
In both cases $\Delta_v$ is the virial overdensity relative to the
critical density of the universe $\rho_c$ given as a fitting form for
a flat universe by \citet{kuhlen2005}, extended to include a time-evolving dark energy:
\begin{equation}
\Delta_v = 18\pi^2\Omega_m(z)[1+a\Theta^b(z)]
\end{equation}
where $\Theta(z) = \Omega_m^{-1}(z)-1$, $\Omega_m(z)$ is the matter
density of the universe as a function of redshift $z$, $a =
0.432-2.001(|w(z)|^{0.234}-1)$, $b = 0.929-0.222(|w(z)|^{0.727}-1)$, and $w$
comes from the dark energy equation of state $w =
w_0+w_a\frac{z}{1+z}$ \citep{chevallier2001,linder2003}.  The gas is assumed to trace the DM
distribution for a given cosmology with a constant gas fraction
$f_g\sim 0.8$,
\begin{equation}
\rho_g(r) = \frac{\delta_c\rho_c }{(r/R_s)(1+r/R_s)^2}f_g f_b
\end{equation}
where $f_b = \Omega_b/\Omega_m$ is the mean baryon fraction of the
universe.

The corresponding ICM pressure and temperature profiles are
computed by integrating the equations of hydrostatic equilibrium and
mass conservation for an ideal gas,
\begin{equation}
\frac{dP}{dr} = \rho_g g(r)
\end{equation}
\begin{equation}
\frac{dM_g}{dr} = 4\pi r^2 \rho_g
\end{equation}
\begin{equation}
P = n k_B T = \frac{\rho_g k_B T}{\mu m_p}
\end{equation}
where $k_B$ is the Boltzmann constant, $m_p$ is the mass of a proton,
$\mu = 0.59$ is the mean molecular weight, and
\begin{equation}
g(r) = -\frac{G(M_{DM}+M_g)}{r^2}
\end{equation}
is the gravitational acceleration including the self-gravity of the
gas.  Once $P(r)$ and $T(r)$ are determined, the entropy
profile $K(r)$ is computed according to
\begin{equation}
\label{eq:entropy}
K = T n^{-2/3}
\end{equation}
where $n$ is the total gas\footnote{This in contrast to typical
observational definitions which use the electron density $n_e$}
particle number density $n = P/k_BT$.  The outer boundary condition is
taken to be a pressure resisting the infall of baryonic matter at the
virial radius; $P_{vir} = f_b \rho_{DM} v_{ff}^2/3$, with
$v_{ff}$ the free--fall velocity at the virial radius as in \citet{voit02}.

\subsection{The Modified Distribution}

After the unmodified entropy distribution has been calculated, we
follow \citet{voit02}, and increase it by a constant $K_0$ as
\begin{equation}
\hat{K}(r) = K(r) + K_0
\end{equation}
where $K(r)$ is the unmodified distribution and $K_0$ is a constant
of order 100 keV cm$^2$,
assumed here to evolve with redshift as
$K_0=K_0(z=0)(1+z)^{\alpha_K}$.  This mimics the preheating of the
ICM, whereby some mechanism had injected a fixed amount of entropy
into the gas prior to its collapse at redshift $z$
\citep{ponman1999,voit02}.

We change variables and re--integrate the equations of hydrostatic
equilibrium using the modified entropy distribution $\hat{K}$.  The
equation of hydrostatic equilibrium then reads
\begin{equation}
\frac{dP}{dr}  = g(r) \rho_g(P,\hat{K}) 
\end{equation}
where, casting the pressure and temperature in terms of $P$ and $\hat{K}$
\begin{equation}
\rho_g = \mu m_p \left[\frac{P}{k_B \hat{K}(M_g)}\right]^{3/5}.
\end{equation}
and
\begin{equation}
k_B T = \hat{K}(M_g)^{3/5}P^{2/5}
\end{equation}

We use the same pressure matching boundary condition as before.  This
procedure results in modified profiles for $P$, $\rho_g$, and $T$ that
can be used to make predictions for different bulk properties of the
cluster, and to establish the expected mass--observable relations.

\subsection{Radial Profiles}

\begin{figure}
\plotone{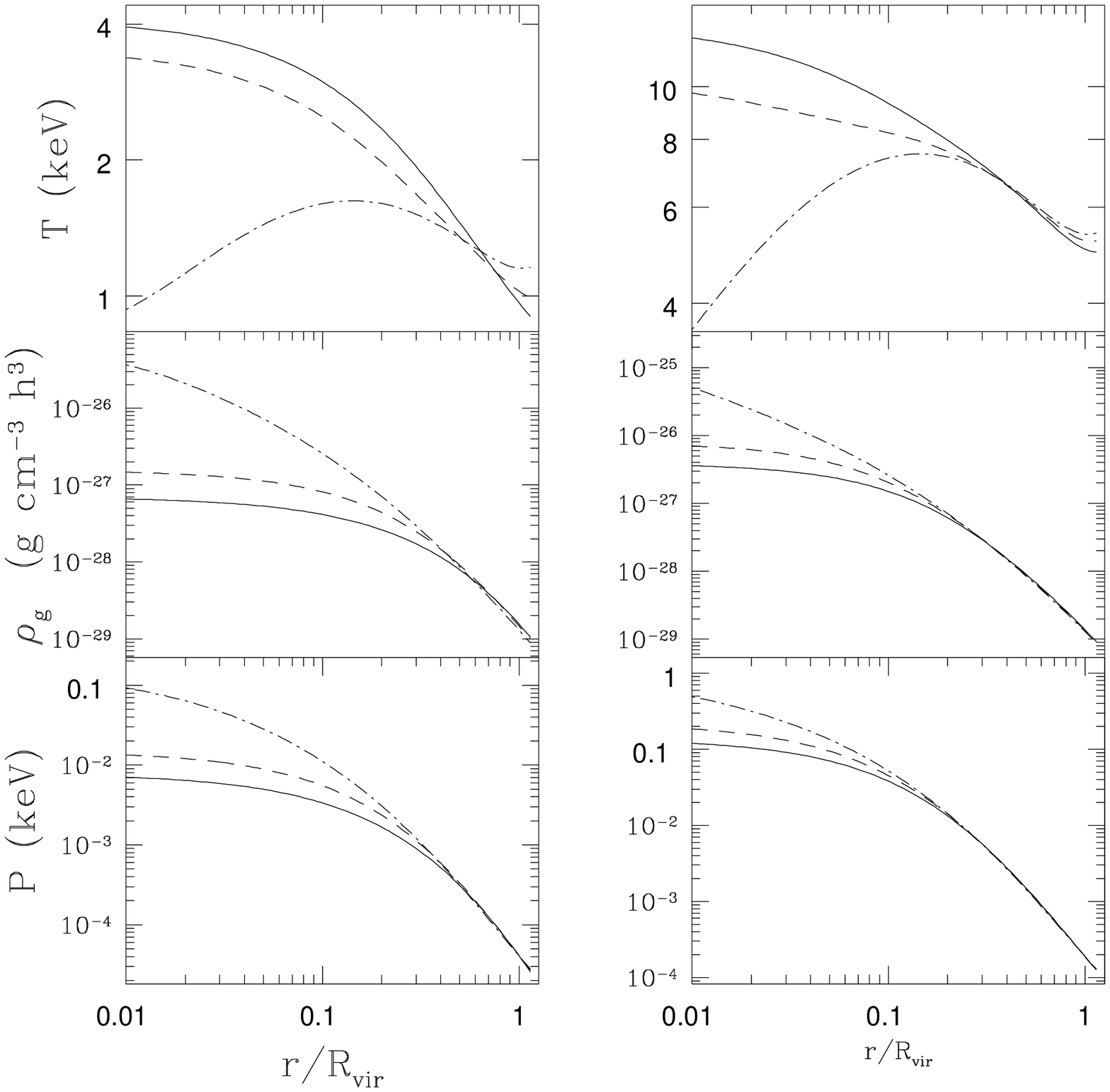}
\caption{The pressure, density, and temperature profiles for a cluster
with a total (gas + DM) mass of $10^{14}$ $h^{-1} M_\odot$ (left
panels) and $10^{15}$ $h^{-1} M_\odot$ (right panels). The profiles
are computed at redshift $z=0$, assuming that the intra--cluster gas
is in hydrostatic equilibrium in an NFW halo.  The three curves in
each panel correspond to different amounts of preheating: $K_0 = 0$
(no preheating; dash--dotted curves), $K_0 = 50$ (dashed curves), and
100 keV cm$^2$ (solid curves; close to our fiducial choice of
125$h^{1/3}$ keV cm$^2$ ).}
\label{fig:profiles}
\end{figure}

Before considering the effects of entropy modification on the bulk
properties of the ICM, it is instructive to visualize its impact on
the radial profiles of clusters of different masses.  In
particular, for reference, we examine the effects of entropy
modifications with $K_0 =50$ and $100$ keV cm$^2$ as compared to the
unmodified model on two clusters with $M_{vir}=10^{14}$ and
$10^{15} h^{-1} M_{\odot}$ respectively, at redshift $z=0$, with our
fiducial $\Lambda$CDM cosmology (see definition below).  The radial
temperature, pressure, and gas density ($T,P,\rho_g$) profiles are
plotted in Figure~\ref{fig:profiles}.

The overall impact of the entropy injection is to depress and flatten
the core pressure and density.  This results in a flattened inner gas
density profile \citep{sun2003,pointecouteau2004} and an outer ($r >
0.1 R_{v}$) temperature profile \citep{vikhlinin2004}, resembling the
results of recent X-ray observations.  It is important to note that the effects of preheating are more pronounced for a low
mass cluster than for a high mass cluster, with the limit that at very
high mass the entropy injection becomes unimportant.  This is because
the injection is a decreasing percentage of the unmodified entropy at
a given radius as the cluster mass increases.

\subsection{Observables}

In order to judge the fitness of our model for use in later
cosmological measurements, we first need to confirm that it can
reproduce the observed scalings for the $M-T$ and $L_x-T$ relations,
given the right choices for the input parameters $f_g$ and $K_0$.  We
recall that the X--ray luminosity scales approximately as $L \propto \rho_g^2$, and
define the luminosity weighted temperature as
\begin{equation}
T_{\rm lum} =  \left(\int \rho_g^2(r) T(r) dV\right) \left(\int \rho_g^2(r) dV\right)^{-1}
\end{equation}
where the integrations (as in all the following cases) are performed
over the entire volume of the cluster out to the virial radius
$r=R_v$.

We then assume a fully ionized H--He plasma with a hydrogen mass
fraction $X=0.75$, $n_e = \rho_g (\frac{X}{m_p} +
\frac{1-X}{2m_p})$ and $n_H = \rho_g \frac{X}{m_p}$, and calculate the X--ray
luminosity as
\begin{equation}
L_x = \int \int n_e n_H \Lambda_\nu(T) dVd\nu.
\end{equation}
Here $\Lambda_\nu(T)$ is the \citet{raymondsmith77} cooling function
for a gas of metalicity $Z = 0.3Z_{\odot}$ for either the bolometric
or the K-corrected soft X--ray emissivity in the $0.5-2.0$ keV
band.

Inverse Compton scattering by electrons in the hot ICM
manifests itself as a fractional change in the temperature of the
Cosmic Microwave Background (CMB), which at frequencies less than $\nu
\sim 218$ GHz appears as a decrement \citep{sunyaev1972,sunyaev1980}.
This effect, often referred to as the thermal SZ effect, provides a
useful probe of the electron pressure $P_e$ integrated along the
line--of--sight (assuming the electrons and gas are in thermal
equilibrium), which complements the X--ray luminosity measurements.
We calculate both the central and integrated SZ flux according to the
prescription in \citet{mccarthy2003a,mccarthy2003b}.

The amplitude of the SZ effect is encoded in the Compton parameter, 
\begin{equation}
y(\theta) = \frac{\sigma_T}{m_ec^2}\int P_e dl
\end{equation}
where $\sigma_T$ is the Thompson cross section and $P_e=n_e k_B T$ is
the electron pressure integrated along the line of sight $dl$.  The
central decrement $y_0=y(\theta=0)$, and the SZ flux is
\begin{equation}
S_\nu = j_\nu(x) y_{int}
\end{equation}
where $x=h\nu/k_B T_{CMB}$ with $T_{CMB} = 2.728$ \citep{fixsen1996}
is the dimensionless frequency, $j_\nu(x) = 2(k_B T_{CMB})^3 (hc)^{-2}
f_\nu(x)$ describes the spectrum of the SZ effect, and
\begin{equation}
y_{int} (\leq \theta) = 2\pi \int_{0}^{\theta} y(\theta ') \theta ' d\theta '
\end{equation}
We choose our observable as the integrated SZ flux,
\begin{equation}
S_\nu = y_{int} \frac{2 (k_B T_{CMB})^3}{(hc)^2}f_\nu,
\end{equation}
evaluated at the frequency $\nu = 150$ GHz.

\section{The Fiducial Model and Cosmology}
\label{sec:fiducial}

In this section, we discuss the parameter choices in our fiducial
model.  Our aim is to demonstrate that our fiducial model is in rough
agreement with a number of existing observations.

The observations we are mainly interested in matching are the scaling
relations between cluster mass and observables, such as
$M_{200}-L_{bol}$, $M_{500}-S_\nu/f_\nu$, and $L_{bol}-S_\nu/f_\nu$,
where $M_{\Delta}$ represents the mass within a spherical overdensity
of $\Delta \rho_c$.  These relations drive the constraints we will
derive below, and they have been estimated in observations over a range of
redshifts. In comparisons to published data, we obtain the relevant
cluster mass assuming an NFW profile with a concentration parameter
$c = 5$. We find that the data suggest a non--evolving
($\alpha_K = 0$) entropy floor of $K_0 = 125 h^{-1/3}$ keV cm$^2$ with
a fixed gas fraction $f_g = 0.8$, which are roughly consistent with the observations of
\citet{ponman1999} and \citet{lloyddavies2000}.

\begin{figure}
\plotone{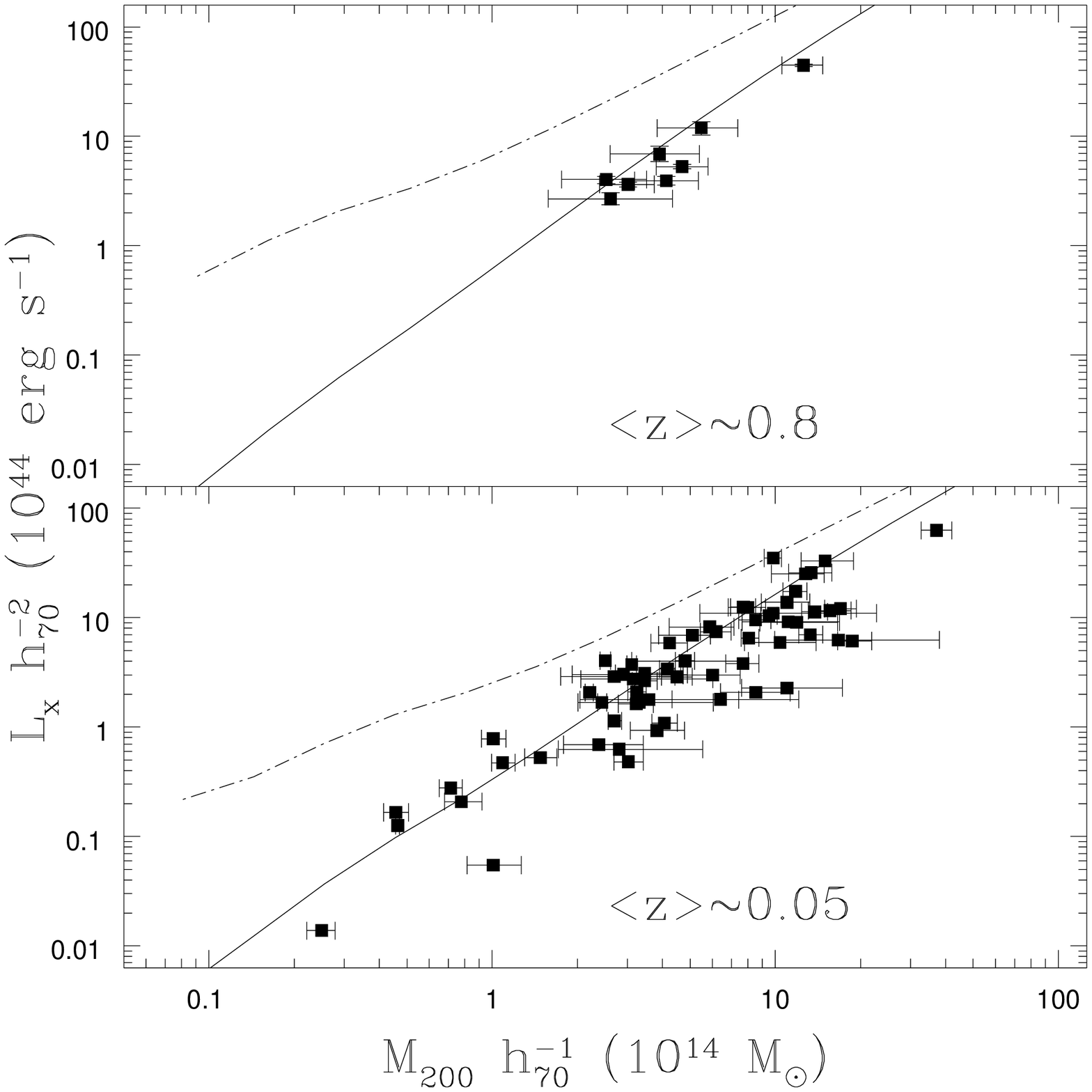}
\caption{The local (bottom panel) and high--redshift (top panel)
mass-luminosity relation between mass and X--ray luminosity for the
self--similar model without preheating (dot--dashed curves) and the
fiducial model, with a non-evolving entropy floor $K_0 = 125 h^{1/3}$
keV cm$^2$ (solid curves). The data points show measurements in the
range $0<z<0.1$ \citep{reiprich2002} and $0.7<z<0.9$
\citep{maughan2005}.  The model predictions are computed at
corresponding mean redshifts of $\langle z\rangle = 0.05$ and $\langle
z\rangle = 0.8$. }
\label{fig:massluminosity}
\end{figure}

\begin{figure}
\plotone{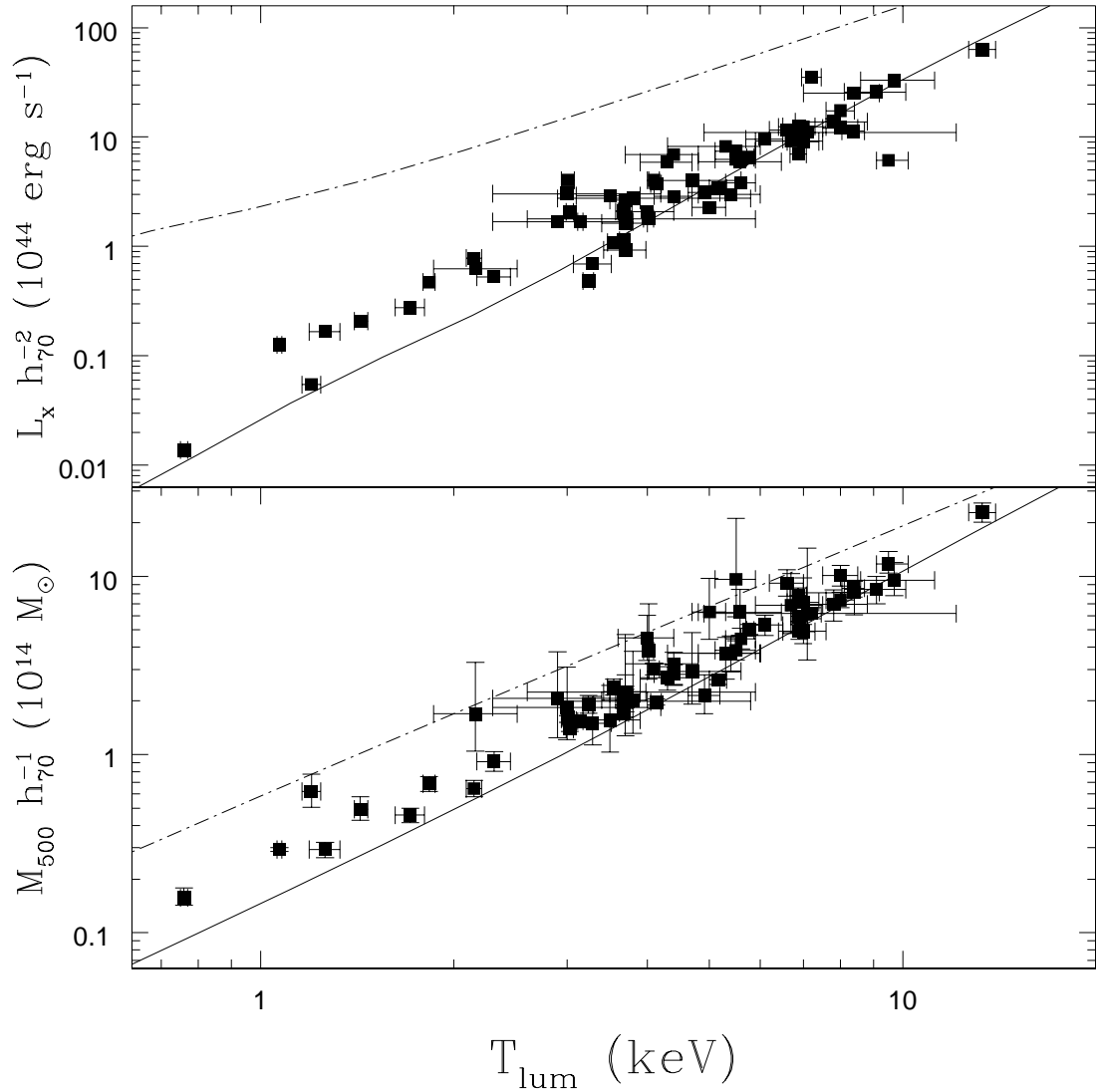}
\caption{The local mass--temperature (bottom panel) and
temperature--luminosity (top panel) relations for the self--similar
model without preheating (dot--dashed curves) and the fiducial model
with a non-evolving entropy floor $K_0 = 125 h^{1/3}$ keV cm$^2$ (solid
curves). The data points are from local measurements at $0<z<0.1$
\citep{reiprich2002}, and the model predictions are computed at the
corresponding mean redshift of $\langle z\rangle = 0.05$.}
\label{fig:localscaling}
\end{figure}

In Figure~\ref{fig:massluminosity}, in the lower panel, we show the
local $M_{200}-L_{bol}$ scaling relation found by \citet{reiprich2002}
at $0<z\lsim 0.1$, along with the predictions from the unmodified
model (dot--dashed curve) and from our fiducial model (solid curve) at
$z = 0.05$.  In the upper panel of the same figure, we show the
observational data of \citet{maughan2005} for the same scaling
relation over the redshift range $0.7<z<0.9$, compared to our
predictions for both the unmodified and the fiducial models at $z =
0.8$.  In the model predictions in both panels, we adopt the same
cosmology as used in the observational papers (different from our
fiducial cosmology, discussed below).
The plot shows good agreement between the data and our fiducial model
over the full redshift range (and the unmodified self-similar model is
clearly ruled out).  The models also predict the ICM gas temperature,
and as another check, in Figure~\ref{fig:localscaling} we compare
these predictions to data on the local $T-M_{500}$ and $T-L_{bol}$
scalings from \citet{reiprich2002}. Again, we find rough agreement
between our predictions and the observed data, although the models
slightly overpredict the temperatures of the lowest--mass clusters.
We find that our models are also in good agreement with other
determinations of both the local mass--temperature
\citep{white1997,girardi1998,finoguenov01,sanderson2003} and
temperature--luminosity
\citep{markevich1998,arnaud1999,fairley2000,novicki2002} relations.

\begin{figure}
\plotone{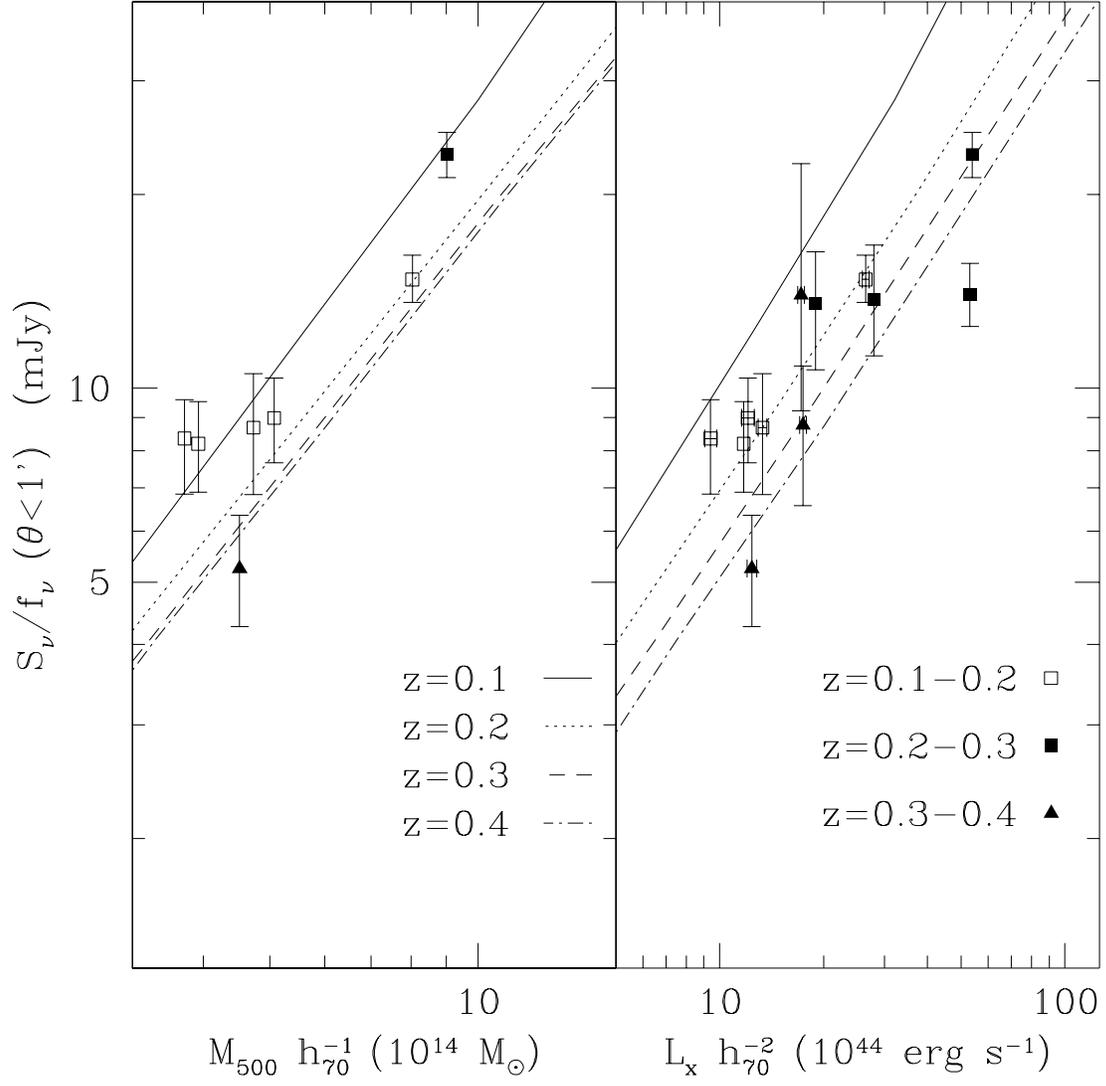}
\caption{Model predictions at four different redshifts in the range
$0<z<0.4$ for scaling relations involving the SZ flux, in the fiducial
model with a non--evolving entropy floor of $K_0 = 125 h^{1/3}$ keV
cm$^2$.  The data points with errors, shown in three different
redshift bins, are taken from \citet{mccarthy2003a} and references
therein.}
\label{fig:szscaling}
\end{figure}

The predictions of the fiducial model were then compared to
measurements of the SZ decrement.  In Figure~\ref{fig:szscaling}, we
show our model predictions at four different redshifts between
$0.1<z<0.4$, for both the $M_{500}-S_{arc,\nu}/f_\nu$ and
$L_{bol}-S_{arc,\nu}/f_\nu$ scalings, where $S_{arc,\nu}/f_\nu$ is the
frequency independent integrated SZ flux in a circle with radius 1 arc-second centered on
the cluster.
Also shown in the figure are the observational data from \citet{mccarthy2003b} and references therein, for the full redshift range $z \sim 0.1-0.4$.  Again, we find good agreement
between our fiducial model and the observed scalings involving the SZ
decrement -- although the data is still sparse, and shows a large
scatter.

\begin{figure}
\plotone{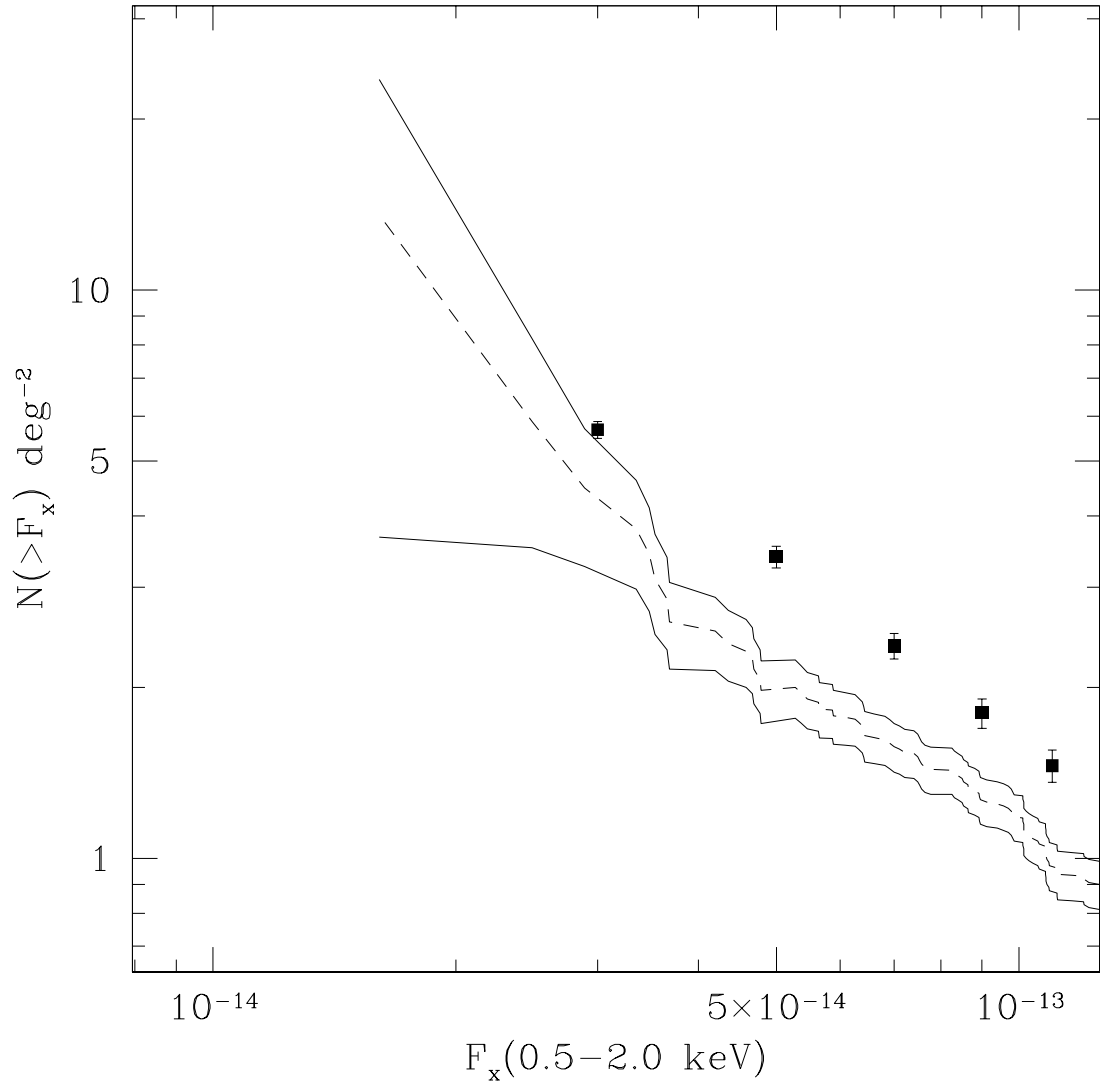}
\caption{The cumulative X-ray source counts predicted in our fiducial
model, as a function of soft X--ray flux in the 0.5-2.0 keV band,
compared to observations by \citet{vikhlinin1998} (dashed curve) bracketed
by corresponding $1\sigma$ error bars (solid curves).  
The fiducial model
predictions are shown as filled squares with Poisson error bars.
At the flux limit of $F_x>3\times 10^{-14}$ erg s$^{-1}$ used in our
analysis, we find a source density of 5.5 deg$^{-2}$, in agreement
with observations, while at higher fluxes, the model somewhat
overpredicts the number of clusters.}
\label{fig:logNlogS}
\end{figure}

For the fiducial cosmology, we adopt $(\Omega_m h^2, \Omega_{DE}, w_0,
w_a, \sigma_8, \Omega_b h^2, n_s)= (0.14, 0.73, -1.0, 0, 0.7, 0.024,
1.0)$.  With the exception of $\sigma_8$, these values are taken from
the best-fit $\Lambda$CDM model found in the first--year {\it WMAP}
data \citep{spergel2003}. It has been shown previously that the local
cluster mass--temperature relation implies a lower value of $\sigma_8$
than CMB observations \citep{seljak2002}.  We arrive at the same
conclusion, when we adjust $\sigma_8$ such that the model predictions
agree with the number counts of X-ray clusters.  For the flux limit of
$F_x(0.5-2.0\textrm{ keV}) > 3\times 10^{-14}$ erg s$^{-1}$ cm$^{-2}$ of the deepest existing
X--ray surveys \citep{vikhlinin1998,gioia2001,rosati2002}, which is
also approximately the limiting flux proposed for a new all--sky
survey \citep{haiman2005}, we find that we need to set $\sigma_8 =
0.7$ in order to be consistent with the observed cumulative number
counts (about 5.5 clusters/degree$^2$ for the above threshold; see
Fig. \ref{fig:logNlogS}).  This low value of $\sigma_8$ is, in fact,
consistent with the recent year--three {\it WMAP} results
\citep{spergel2005}, and also with previous estimates based on the
local mass--temperature relation \citep{seljak2002,pierpaoli2003},
galaxy velocity fields \citep{willick1998}, and weak lensing
measurements \citep{brown2002, hoekstra2002,jarvis2003}.  Our fiducial
cosmology is also consistent with other cluster count measurements
using optical \citep{bahcall2003} and X-ray selected samples
\citep{borgani2001,shuecker2003,gladders2006}.

Finally, despite its rough success, we call attention to potentially
interesting tension between our fiducial model predictions and
observations.  In particular, for both the local and high redshift
scaling relations shown in Figure~\ref{fig:localscaling}, the cluster
temperatures are somewhat overpredicted for the lowest mass clusters.
Likewise, Figure \ref{fig:logNlogS} shows that our model
$\log{N}-\log{S}$ relation somewhat overpredicts the number of bright
clusters.  That the current cluster data reveals such tension
highlights the potential of larger surveys in calibrating models of
the ICM simultaneously with cosmology. In the present paper, however,
our goal is to explore a methodology, rather than a specific cluster
model.  We made use of the existing data only to guide us in choosing
a reasonable set of fiducial parameters, and we postpone a more
rigorous study of the viability of this model to future work.

\section{The Phenomenological Model}
\label{sec:selfcal}

Our aim in this paper is to quantify the merits of using a model for
the cluster, as opposed to directly self--calibrating the
mass--observable relation, as proposed in previous studies
\citep{majumdar2004,wang2004}.  The later approach will be hereafter referred to as a ``Phenomenological Model."  To enable a comparison between the
two approaches, we here consider joint cosmological and ``cluster
model'' constraints, adopting simple parametric power--law scalings
for the mass--observable relation to include the latter.  We follow
\citet{majumdar2004} and \citet{wang2004}, and parametrize the X--ray
luminosity as
\begin{equation}
L_{x,bol} = 4\pi F_x d_L^2 = A_x M_{200,15}^{\beta_x} E^2(z) (1+z)^{\gamma_x}
\label{eq:Lx}
\end{equation}
where $d_L$ is the luminosity distance in Mpc, $L_x$ is the
bolometric X--ray luminosity in units of erg s$^{-1}$, $F_x$ is the
observed bolometric X--ray flux in units of erg s$^{-1}$ cm$^{-2}$,
$M_{200,15}$ is the mass of the cluster within an overdensity of
$200\rho_c$ in units of $10^{15}$ $M_\odot$, with a normalization $\log{A_x} = -3.56$ (see \S 5.1 for details).  We apply a K--correction
to obtain the flux in the 0.5-2.0 keV band, $L_x(0.5-2.0 \textrm{keV})
= L_{x,bol} k(z)$, with
\begin{equation}
k(z) = \frac{\int_{\nu_1(1+z)}^{\nu_2(1+z)} \Lambda_\nu(T)
d\nu}{\int_0^\infty \Lambda_\nu(T) d\nu},
\end{equation}
where $\Lambda_\nu(T)$ is the \citet{raymondsmith77} cooling function
for a gas of metalicity $Z = 0.3Z_{\odot}$ and temperature $T$. The
ICM gas temperature in the direct self--calibration case is needed
only for this K-correction, and was estimated using the observed
mass--temperature relation of \citet{finoguenov01},
\begin{equation}
M_{180} = (1\pm 0.2) \times 10^{15} h^{-1} M_\odot \left
(\frac{T}{\beta_T \Delta_v^{1/3} \textrm{ keV}}\right)^{3/2}
\end{equation}
where $M_{180}$ is the cluster mass within an overdensity of 180
relative to the background, $\beta_T = 1.75\pm 0.25$, and $\Delta_v$
is the virial overdensity.

The SZ flux is parametrized similarly as a power--law,
\begin{equation}
S_\nu d_A^2 = A_{sz} M_{200,15}^{\beta_{sz}} E^{2/3}(z)
(1+z)^{\gamma_{sz}}
\label{eq:Snu}
\end{equation}
where $d_A$ is the angular diameter distance in Mpc, and $S_\nu$ has
units of mJy at a frequency of $\nu = 150$ GHz, with a normalization $\log{A_{sz}} = 8.29$ (see \S 5.2 for details).

\section{Simulating Galaxy Cluster Surveys}
\label{sec:surveys}

To simulate the abundance of clusters of galaxies, we begin with the
standard formalism \citep{press1974},
\begin{equation}
\frac{dn}{d\ln{M}} = f(M) \frac{\rho_0}{M} \frac{d \log{\sigma^{-1}}}{d \ln{M}}
\end{equation}
where $M$ is measured out to an spherical overdensity of 180 relative
to the background, $\rho_0$ is the present day background matter
density, $f(M)$ is taken from the extended Press-Schechter fit of
\citet{jenkins2001}, and $\sigma$ is the variance of the linear
density field at redshift $z$.  The power spectrum was calculated
using the fitting formulae of \citet{eisenstein1999}, modified to
include the effects of a time-varying dark energy parameter with $w(z)
= w_0+w_a\frac{z}{1+z}$ \citep{linder2003}.

The quantity we will use below to derive cosmological constraints is
the number of clusters within a bin corresponding to a range of the
observable $\Psi$. We follow \citet{lima2005} and include a constant
log--normal scatter in the cluster mass at a fixed observable, with
standard deviation $\sigma_{\log{M}|\Psi}$.  In the absence of
scatter, we use our mass--observable relation -- either the phenomenological or physical model --
to map the flux bin $\Psi_i < \Psi < \Psi_{i+1}$ to a mass bin
$M_{obs}^i < M_{obs} < M_{obs}^{i+1}$, to compute the expected counts.
In the presence of scatter, for a redshift bin centered at $z_j$ of
width $\Delta z$, the expected counts within $\Psi_i < \Psi <
\Psi_{i+1}$ is modified to
\begin{equation}
N_{i}(z_j) = \Delta \Omega \Delta z \frac{dV}{dzd\Omega} \int
\frac{d\ln{M}}{2} \frac{dn}{d\ln{M}} \left
[\textrm{erfc}(x_i)-\textrm{erfc}(x_{i+1}) \right ]
\end{equation}
where
\begin{equation}
x_i = \frac{\log{M} -
\log{M_{obs}^{i}}}{\sqrt{2\sigma^2_{\log{M}|\Psi}}}
\end{equation}
and $dV/dzd\Omega$ is the comoving volume element, and erfc is the complementary error function.

\subsection{Survey Parameters}
\label{subsec:params}

Our mock X--ray survey has a flux detection threshold of
$F_x(0.5-2.0\textrm{ keV}) = 3\times 10^{-14}$ erg cm$^2$ s$^{-1}$,
comparable to the limit reached in existing deep surveys, as well as
the threshold in the all--sky survey proposed to the Dark Energy Task
Force \citep{haiman2005}.  In our fiducial model, this yields a
cluster surface density of $\sim 5.5$ deg$^{-2}$.  Including scatter
in the mass-observable relation increases the total number of clusters
by $\sim 10\%$ due to preferential upscattering of a larger population
of less massive clusters, particularly at high redshift as the mass
function steepens.  We furthermore impose a strict luminosity floor of
$L_{fl} = 3\times 10^{42}$ erg s$^{-1}$ in the observed band, roughly
corresponding to the luminosity of small groups \citep{arnaud1999},
below which both observations and our physical model become
unreliable.  The logarithmic slope of the fiducial mass vs.  X--ray
luminosity relation is set to $\beta_x = 1.807$ to match the observed
scaling, with no evolution $\gamma_x = 0$
\citep{wang2004}, and a normalization $\log{A_x} = -3.56$. This then
also matches the cluster surface density predicted in the fiducial
physical model, and facilitates a fair comparison of the two
approaches.

The mock SZ survey is modeled after upcoming observations by the South
Pole Telescope \citep{ruhl2003,wang2004}, with a limiting SZ flux of
$S_\nu = 3.0$ mJy at $\nu = 150$ GHz.  This flux threshold produces a
yield that again matches the cluster counts of 5.5 deg$^{-2}$ in the
mock X--ray survey, allowing an easy comparison.  The fiducial
power-law parameters are chosen - as before - to match the local
scalings with no evolution ($\beta_{sz} = 1.68$ with $\gamma_{sz} =
0$), and the normalization is chosen to agree with the number of
clusters predicted by the physical model ($\log{A_{sz}} = 8.29$).

\begin{figure}
\plotone{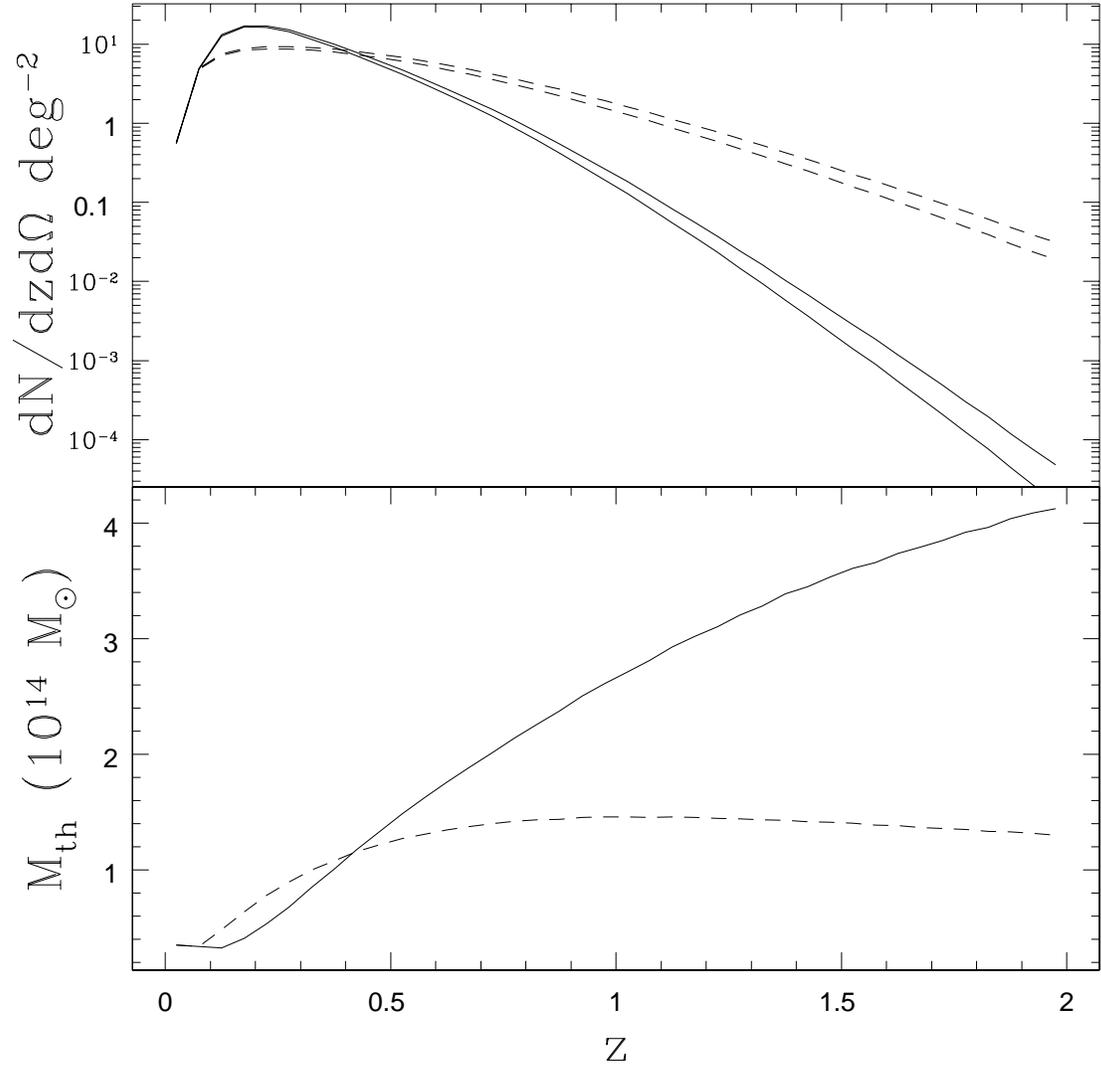}
\caption{The lower panel shows the evolution of the limiting mass
$M_{th}$ corresponding to the flux detection threshold, and the upper
panel shows the number counts of clusters above this mass, in the mock
X--ray (solid curves) and SZ (dashed curves) surveys.  The pairs of
curves in the top panel correspond to the case of no scatter in the
mass-observable relation (lower curve) and a log--normal scatter in
the mass, at fixed observable, of $\sigma_{\log{M}|\Psi} = 0.1$ (upper
curve).}
\label{fig:dNdz}
\end{figure}

In both surveys, and for both the physical and phenomenological model, 
we consider three cases: (1) a single flux
bin extending from the detection threshold up to arbitrarily high
fluxes, (2) 20 flux bins with no scatter, and (3) 20 flux bins with a
log--normal scatter that has a fiducial value of
$\sigma_{\log{M}|\Psi} = 0.1$ for both the X--ray and SZ
mass-observable relation
\citep{lima2005}, but is assumed to be a free parameter.  For each of 
these, we consider two different descriptions of the entropy injection
history.  First, we fit a power-law evolution of the entropy floor of
the form $K_0(z) = K_0(z=0)(1+z)^{\alpha_K}$, with $K_0$ and
$\alpha_K$ as free parameters.  Second, we consider arbitrary
evolution of the entropy floor $K_0$, in which $K_0$ is fit
independently in each of 40 redshift bins of width $\Delta z=0.05$.
In both of these cases, the fiducial entropy floor is set to $K_0 =
125 h^{1/3}$ keV cm$^2$ with no evolution.  For reference, in
Figure~\ref{fig:dNdz} we show the redshift evolution of $dN/dzd\Omega$
and the of the mass $M_{th}$ corresponding to the detection threshold
in both the X--ray and SZ surveys, each for the case of no scatter and
$\sigma_{\log{M}|\Psi} = 0.1$.  Finally, for the purpose of
comparison, the sky coverage in both surveys is taken to be
$\Delta\Omega = 4000$ deg$^2$, the size planned for SPT
\citep{ruhl2003}. For surveys covering a  different
solid angle, the constraints we obtain below scale as $\propto
\Delta\Omega^{-1/2}$.

\subsection{CMB Anisotropies}

Our simulated CMB survey is modeled after a near to medium--term
space based all-sky CMB survey, similar to the proposed Planck
mission with bands at 100, 143, and 217 GHz.  We assume fractional sky
coverage of $f_{sky} \approx 0.8$, and perfect subtraction of
foregrounds.  The $C_\ell$ coefficients are calculated up to
$\ell_{max} = 2\times10^{3}$ using KINKFAST \citep{corasaniti2004}, a
version of CMBFAST \citep{seljak1996} modified to include a time
varying $w$.  See, e.g., \citet{wang2004} for more details.

\section{The Fisher Matrix Forecasts}
\label{sec:fisher}

\subsection{Background}

With a fiducial cosmology and a prescription for simulating galaxy
surveys, one can compute lower limits on statistical errors achievable
on model and cosmological parameters.  In our self--calibrating
approach, the cluster observations are used to simultaneously
constrain both the cosmology and model parameters.

Assuming a reasonably well--behaved likelihood function $\cal{L}$, the
Fisher matrix is a quick method to forecast joint parameter
uncertainties in a multi--parameter fit \citep{tegmark1997}.  The
Fisher matrix is defined as
\begin{equation}
F_{\alpha\beta} =  \left <-\frac{\partial^2 \log{\cal{L}}}{\partial p_\alpha \partial p_\beta} \right >
\end{equation}
where $p_\alpha$ and $p_\beta$ are free parameters.  The inverse of
the Fisher matrix gives the best attainable
covariance matrix $C$.  Therefore, the constraints on any parameter
$p_\delta$, marginalized over all other parameters, is
\begin{equation}
\Delta p_\delta^2  = C_{\delta\delta} = \left [F^{-1}\right ]_{\delta\delta}
\end{equation}
We further define a degeneracy coefficient $\xi_\delta$, which
quantifies the degree to which degeneracies among the parameters increase
the marginalized parameter error over the single--parameter error,
\begin{equation}
\xi_\delta^2 =C_{\delta\delta} F_{\delta\delta}.
\end{equation}
The degeneracy coefficient $\xi_\delta$ ranges from $1$ for a purely
non-degenerate matrix, to $\infty$ for fully degenerate parameters.

The combined Fisher matrix for a series of $N$ separate experiments is
the sum of the individual Fisher matrices, i.e.
\begin{equation}
F_{tot} = \sum_{i = 1}^{N} F_{i}
\end{equation}
which motivates the definition of a synergy coefficient
$\zeta_\delta$, to quantify the fractional improvement in the
marginalized error on a parameter $p_\delta$ after combining
experiments versus adding the individual experiment errors in
quadrature, as
\begin{equation}
\zeta_\delta^2 = C_{\delta\delta} \sum_{i = 1}^{N} \left ([C_i]_{\delta\delta} \right )^{-1}
\end{equation}
The synergy coefficient ranges from $1$ for no improvement from
combining the experiments beyond adding the individual constraints
in quadrature, to $0$ if the degeneracies are fully broken, in the
limiting case that the combined Fisher matrix delivers a perfect
knowledge of $p_\delta$.

\subsection{Galaxy Cluster Counts}

The Fisher matrix for a mock cluster survey, $F_{\alpha\beta}^c$,
can be calculated from the number counts of clusters as a function of
the observable $\Psi$.  We define the counts Fisher matrix as
\begin{equation}
F_{\alpha\beta}^{c} = \sum_{i} \frac{\partial N_i}{\partial
p_\alpha}\frac{\partial N_i}{\partial p_\beta}\frac{1}{N_i}
\end{equation}
where $i$ is the bin index, and the error is purely Poisson counting
error (ignoring the modest increase due to sample variance,
\citet{hukrav2003}.  This represents a sum of several Fisher
matrices, one for each independent flux and/or redshift bin.

Therefore, given redshift information, it is possible to derive
cosmological constraints from a combination of the flux distribution
and redshift evolution of cluster counts.  We bin the data by flux -
as before - and redshift, giving a Fisher matrix
\begin{equation}
F_{\alpha\beta}^{cz} = \sum_{i} \sum_{j} \frac{\partial N_i(z_j)}{\partial p_\alpha}\frac{\partial N_i(z_j)}{\partial p_\beta}\frac{1}{N_i(z_j)}
\end{equation}
where the sum is taken out to a limiting redshift of $z_f$.  The first $n_{bin}-1$ flux bins are spaced evenly in log-space from the limiting flux over five orders of magnitude, with a final bin extending out to $\infty$.  We vary all the cosmological and model parameters, with the exception of $\Omega_b h^2$ and $n_s$, which are later included via CMB observations.  Finally, surveys in different observables are not assumed to overlap.

\subsection{CMB Observations}

In addition to constraints from cluster counts, we calculated the
Fisher matrix for observations of CMB temperature and polarization
anisotropies.  We assume a Planck-like satellite mission\footnote{see
www.rssd.esa.int/index.php?project=Planck.} including both the
temperature $T$ and $E$-mode polarization auto-correlation, as well as
the TE cross-correlation, while neglecting $B$--mode polarization.
The full CMB Fisher matrix is then calculated as in
\citet{zaldarriaga1997}.  Each frequency channel
was assumed to provide independent cosmological constraints.  See
\citet{wang2004} for full details.
 
\section{Results and Discussion}
\label{sec:results}

In order to present our results more clearly, we will proceed in
pedagogical order, considering successively more realistic
assumptions.

\subsection{Cosmology--Only Constraints}

In this subsection, we study the limiting case in which the cluster
model parameters are known perfectly. We begin with the most na\"{i}ve
experiment, using a single flux bin (i.e. only total cluster counts
above the detection threshold, as a function of redshift), and
ignoring any scatter in the mass--observable relations. The
constraints on the cosmology along with their associated degeneracy
coefficients for X--ray and SZ surveys, and the synergy coefficient
for the combination are presented in Table
\ref{table:1bin0scat.cosmo}.
The physical model breaks many cosmological degeneracies, and
therefore gives much better constraints, for the X--ray and SZ surveys
individually as compared to the phenomenological model.  
This is particularly true of the dark energy equation of state
parameters $w_0$ and $w_a$, which improve by a factor of $\sim 2$ in
the X--ray and $\sim 4$ in the SZ survey.  The single parameter
sensitivities -- the diagonal elements of the Fisher matrix -- for
$w_0$ and $w_a$ are comparable in both approaches (physical or
phenomenological model). We find that the degeneracy
breaking in the physical model is due to better absolute sensitivity
to $\Omega_m h^2$ .  This is because changing $\Omega_m h^2$ while
holding $\Omega_b h^2$ fixed (and assuming a flat universe) changes
the mean universal baryon fraction.  Accordingly, the value of the
observables for a cluster of fixed mass will vary with $\Omega_m
h^2$. As a result, changes in $\Omega_m h^2$ manifest themselves
strongly in the redshift distribution of cluster counts in a
flux--limited survey, even at fixed $\Omega_b h^2$, when the physical, rather than phenomenological model, is used to
calculate the threshold mass.  This difference should be robust, in
the sense that any physical model of cluster structure is likely to
predict $\Omega_m h^2$--dependence of the X--ray flux and SZ decrement
(over and above the $\Omega_m h^2$ dependence contained in the Hubble
parameter $E(z)$ in the power--law relations in equations~\ref{eq:Lx}
and \ref{eq:Snu}).

\begin{figure}
\plotone{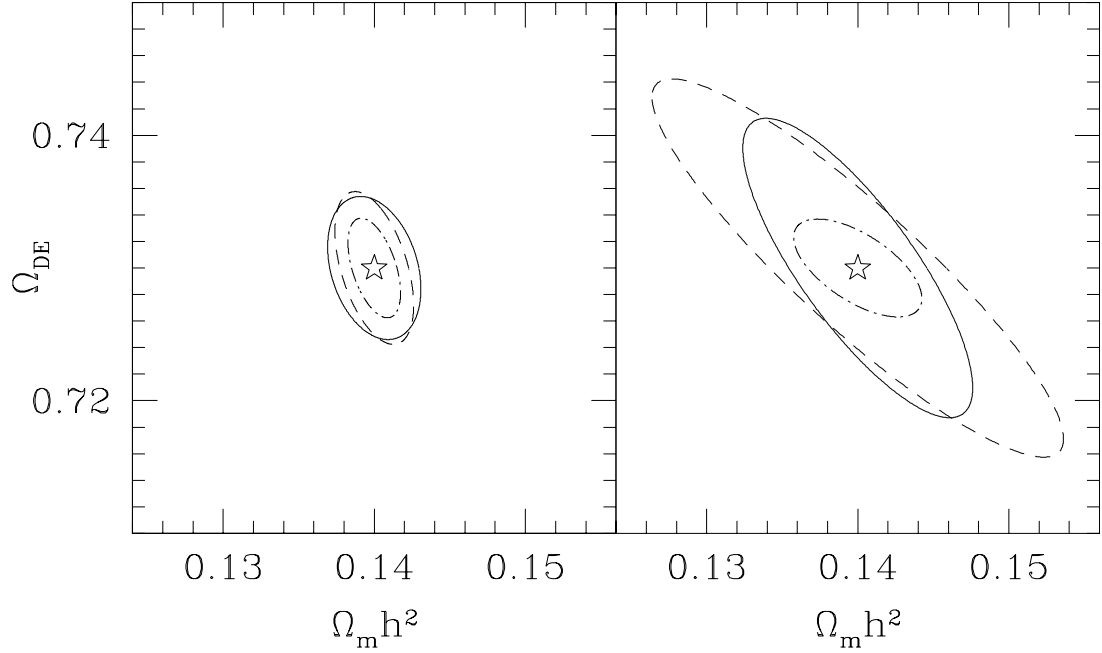}
\caption{Parameter constraints in the $\Omega_mh^2-\Omega_{DE}$ plane
for direct self--calibration of the mass--observable relation (right
panel) and using the physical model (left panel). The constraints have
been marginalized over all other cosmological parameters. The figure
assumes a single flux bin (the shape of the mass function is not used)
and no scatter in the mass-observable relation (see
Table~\ref{table:1bin0scat.cosmo}). In both panels, we show constraints
from the mock X--ray (solid ellipses) and SZ (dashed ellipses) surveys
individually, and from the combination of the two surveys (dot--dashed
ellipses).  Note the overall improvement in the constraints when the
physical model is used, arising from breaking the $w-\Omega_mh^2$
degeneracies within each survey.}
\label{fig:cosmo.only}
\end{figure}

While using the physical model significantly improves the degeneracy
coefficients for cosmological parameters in both the X--ray and SZ survey,
it also removes the synergy between the two, and degrades the synergy
coefficient for their combination.  For example, $\zeta_{w_0} \sim
0.60$ for the phenomenological model, as compared to $\zeta_{w_0} \sim 1$
for the physical model.  We illustrate this effect by showing the
marginalized constraints in the $\Omega_m h^2 - \Omega_{DE}$ plane in
Figure~\ref{fig:cosmo.only}. The error ellipses shown for phenomenological
model (right panel) are larger than for the physical model (left
panel), and they are also more complementary, with major axes pointing
in different directions.  The overall constraints on the cosmology
are, however, are still better when the physical models are used.

We next present constraints using 20 flux bins for both the X--ray and
SZ surveys in Table~\ref{table:10bin0scat.cosmo}.  Using multiple bins
allows additional information to be extracted from the shape of the
cluster mass function.  As a result, the degeneracy coefficients for
all parameters are substantially reduced in both surveys.  In fact, all parameters become nearly non-degenerate, and the synergy coefficients in both surveys using either the phenomenological or physical model are close to $\sim 1.$  Therefore, these constraints reflect the individual parameter sensitivities of the surveys.  These sensitivities are roughly equal in the two approaches for all parameters, with the exception of $\Omega_m h^2$, to which the physical model is a factor of $\sim 6$ more sensitive.  This is not surprising, since the cosmological dependence of the physical model reduces to that of the phenomenological model in the limit of $K_0 \sim 0$, expect for the $\Omega_m h^2$ dependence.  We therefore confirm our interpretation of constraints using 1 flux bin, and illustrates quantitatively the power of the model to resolve $\Omega_m h^2$.  This sensitivity, under more realistic assumptions, will help the physical model break more degeneracies than the phenomenological model.

Constraints for more realistic surveys, in which scatter in the mass-observable relation will degrade the information that can be extracted from the mass--function shape, are summarized in
Table~\ref{table:10bin10scat.cosmo}.  These use 20 flux bins in both
the X--ray and SZ, but also include a log--normal scatter as a free
parameter.  The scatter smears out the shape of the mass function, and
increases the degeneracy coefficients for all parameters and in both
surveys, to values comparable to those using a single flux bin and no scatter.  Overall, the constraints are within a factor of two of the
1--bin, no--scatter case, indicating that scatter destroyed most of
the information from the mass function shape. The phenomenological case
still has higher overall degeneracy coefficients, which makes the
X--ray and SZ constraints more synergistic in combination.  On the
other hand, the absolute constraints for all parameters are better
when the physical model is used.  

We also find that imposing priors on the scatter does not significantly improve constraints on the cosmology.  This contrasts with the results of \citet{lima2005}.  They, however, use a fixed mass threshold.  For a flux limited survey, constraints on the cosmology degrade when a log-normal scatter is introduced as a free parameter, not primarily because of degeneracies among the scatter and the cosmology, but rather because the scatter itself washes out the cosmological sensitivity of the mass--observable relation.  To test this hypothesis, we consider constraints on the cosmology and scatter using a fixed mass threshold versus a fixed flux threshold, both with 1 bin.  We find that when a fixed mass threshold is used, fixing the scatter improves cosmological constraints by a factor of 2 in most cases, and up a factor of 4 in some.  However, when a fixed flux threshold is used, the improvement is typically no more than 30\%.  This implies that the mass--observable relation, which is flattened by the inclusion of scatter, is driving the constraints on the cosmology.

In summary, for the idealized case when the cluster structure or
mass--observable relation is assumed to have no uncertainty, we find
that the physical model leads to better sensitivity to dark energy
parameters in both the SZ and X--ray surveys individually, and also in
the case when the two surveys are combined. We find that this
improvement arises because the physical model predicts an increased
sensitivity of the observables to $\Omega_m h^2$, which removes
degeneracies between $\Omega_m h^2$ and dark energy
parameters. Interestingly, the degeneracies are improved so
substantially in the individual surveys, that the combination of the
two surveys is {\it less} synergistic than in the phenomenological case.

\subsection{Self-Calibrated Constraints}

In this subsection, we present self--calibrated constraints, in which
the parameters of the phenomenological or physical
model, are assumed to be a--priori unknown.  We will focus on the extent
to which this degrades cosmological constraints.  It is worth noting,
however, that such a self-calibration approach can yield tight
constraint not just on the cosmology, but also on the properties of
the clusters themselves.

In Table~\ref{table:1bin0scat.cal}, we show self--calibrated
constraints in the simplest case of 1 flux bin with no scatter.  The
results show that the phenomenological model parameters
are extremely degenerate with the cosmology, suggesting their ability
to introduce changes in $dN/dz$ in ways that mimic the effects of
changing the cosmology.  This has been noted previously: for example,
uncertainties in the normalization of the observed mass--temperature
relation result in large uncertainties in the value of $\sigma_8$
measurements \citep{seljak2002,pierpaoli2003} when this value is
inferred from the cluster temperature function.  The severe
degeneracies between cosmological and mass-observable relation
parameters are manifest in the high values of the degeneracy
parameters, especially for $\sigma_8$, with of $\xi_{\sigma_8}\sim 100$ and
$600$ for the X--ray and SZ surveys respectively.  When the X--ray and
SZ surveys are combined, it is possible to break some of these
degeneracies, but for $\sigma_8$, and for the dark energy parameters
in particular, the synergy coefficients $\zeta_{w_0,w_a} \sim 1$, indicating
that uncertainty in the slope, normalization, or evolution of the
mass--observable relations in the X--ray or SZ can mimic the effects
of variations in these cosmological parameters.

The above contrasts sharply with the results we obtain using a
physical model -- as the bottom half of
Table~\ref{table:1bin0scat.cal} shows,
there is only minor degradation of the cosmological parameter
constraints due to the uncertainties in $f_g$, $K_0$, and $\alpha_K$.
In this case, the X--ray parameter degeneracies on average increase
only by a factor of $\sim 2-3$ as compared to the case of perfect
knowledge of the model parameters, while the SZ degeneracies increase
by no more than $40\%$.  This demonstrates that the physical model
parameters $f_g$, $K_0$, and $\alpha_K$ do not easily mimic the
effects of a change in cosmology.  The comparison with the phenomenological 
model is especially dramatic for the dark
energy equation of state parameters in the SZ survey, whose
constraints are improved by a factor of $\sim 10$.  In addition,
because these parameters simultaneously determine both the X--ray and
SZ mass-observable relations, and because they enter into the
expressions for the flux differently (through $\rho_g^2$ for the
X--ray and $\rho_g$ for the SZ flux), the two surveys are more
complementary than in the case when these parameters are fixed.
Therefore, the synergy coefficients are smaller than their analogues
in Table~\ref{table:1bin0scat.cosmo}.

In Table~\ref{table:10bin0scat.cal}, we extract information from the
shape of the mass function, by using 20 flux bins, while still
assuming no scatter in the mass--observable relation.  As before, we
find that the shape of the mass function helps to break many of the
parameter degeneracies, with $\xi \sim 1-2$ for all parameters, and thus these
constraints again are indicative of the single parameter sensitivities.  
In the middle and the bottom set of rows
in this table, we consider two different parametrizations of the
entropy injection history in the physical model: power--law, and
arbitrary evolution of $K_0$ with redshift.  In both cases, the
degeneracy coefficients for all parameters are significantly lower
than those in the direct self--calibration models.  They are somewhat
higher for arbitrary $K_0$--evolution by a factor of $\sim 1.5$,
indicating that relaxing the assumption of a power--law evolution can
better mimic the effects of dark energy.  However, in this case of
arbitrary $K_0$ evolution, the X--ray and SZ constraints are more
synergistic in combination.  Therefore, when both the X--ray and SZ
results are considered, the absolute parameter constraints are
degraded by no more than $\sim 20-30\%$ even when $K_0$ can evolve
arbitrarily with redshift.  We also see, as in the previous section, far greater sensitivity to $\Omega_m h^2$ when using the physical rather than the phenomenological model.

\begin{figure}
\plotone{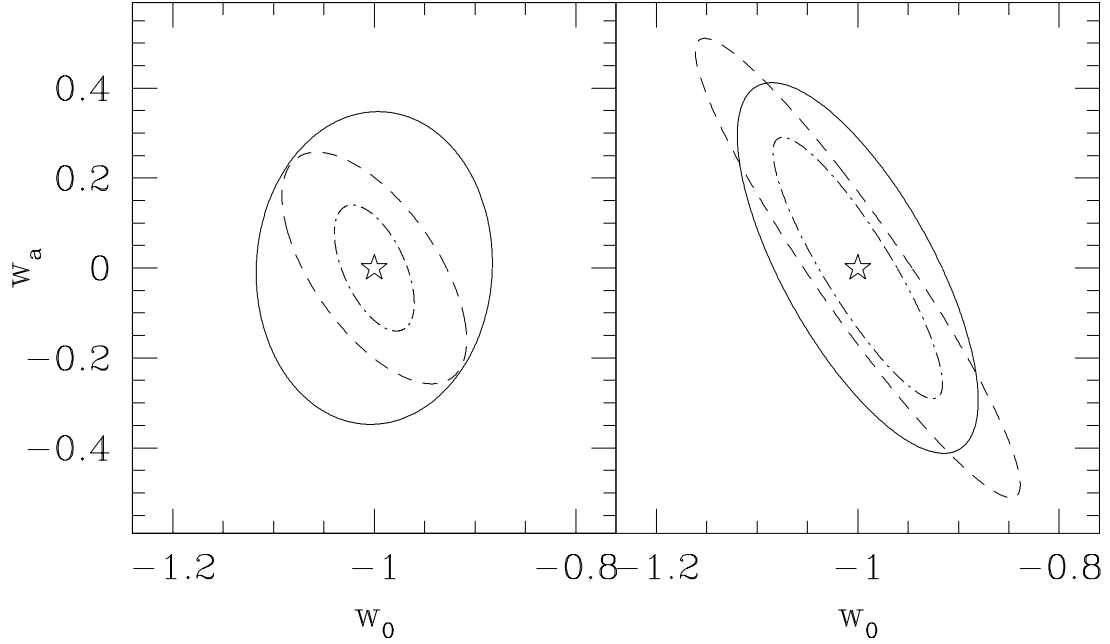}
\caption{Parameter constraints in the $w_0-w_a$ plane for direct
self--calibration of the mass--observable relation (right panel) and
using the physical model (left panel).  The constraints have been
marginalized over all other cosmological parameters. The figure
assumes $n_{bin} = 20$ independent flux bins, and a log--normal
scatter in the mass--observable relation that is aprori unknown (with
the fiducial value of $\sigma_{\log{M}|\Psi} = 0.1$; see
Table~\ref{table:10bin10scat.cal}).  In both panels, we show
constraints from the mock X--ray (solid ellipses) and SZ (dashed
ellipses) surveys individually, and from the combination of the two
surveys (dot--dashed ellipses).  Note the improved degeneracy breaking
in the physical model, when data from the two surveys are combined.}
\label{fig:darkenergy}
\end{figure}

Finally, the constraints in the most realistic among our mock surveys,
which includes 20 flux bins and a log--normal scatter as a free
parameter, is presented in Table~\ref{table:10bin10scat.cal}.  As we
found in the the cosmology--only constraints above, the uncertainty in
the scatter increases the degeneracies for all parameters, in both
surveys, and for both the phenomenological and physical model results.
For the phenomenological model, we recover the
complementarity of the two surveys.  However, both cases of the
physical model give smaller degeneracy coefficients and better
absolute constraints for all parameters.  In particular, using the
physical model improves $w_0$ and $w_a$ constraints by a factor of
$2$ relative to the phenomenological 
model.  We also note that allowing arbitrary
$K_0$--evolution still does not significantly degrade constraints,
even in the presence of uncertain scatter, relative to a power--law
$K_0$--evolution.  Interestingly, we find that this is because for a
physical model with arbitrary $K_0$ evolution, the X--ray and SZ
surveys are exceptionally complementary.  In particular, the synergy
coefficient for the dark energy parameters $\zeta_{w_0,w_a} < 0.7$,
indicating that constraints improve by more than 30\% (over adding the
constraints in quadrature) when the X--ray and SZ surveys are
combined.  This is illustrated visually in
Figure~\ref{fig:darkenergy}, where we show marginalized constraints in
the $w_0-w_a$ plane. The figure shows a significant reduction in the
size of the error ellipse when the two surveys are combined. In
particular, this reduction is much more significant in the physical
model (left panel) than in the case of the phenomenological model (right panel).

\subsection{Complementary Observations}

Finally, we contrast the constraints we find from the mock cluster
surveys to measurement of CMB
anisotropies (by Planck).  The
constraints and degeneracy coefficients we find for Planck individually are listed in
Table~\ref{table:complementary}.  The table shows that the forecasts
for the combined X--ray and SZ cluster surveys, even when we allow for
unknown scatter, and for arbitrary entropy--injection history, compares
favorably with Planck, including $\Omega_m h^2$
which is known to be measured to exquisite precision with the CMB.

We also investigate the utility of CMB measurements in breaking the
parameter degeneracies inherent in the X--ray and SZ cluster surveys.
Constraints for the combination of Planck with the X--ray and SZ
cluster surveys (including 20 bins and uncertain scatter, as in
Table~\ref{table:10bin10scat.cal}) are listed in
Table~\ref{table:10bin10scat.cmb}.  The inclusion of the Planck data
adds substantial sensitivity to the individual parameters, which
results in a significant increase in the degeneracy coefficients.  We
find that when Planck data is assumed to be available, the combination
of X--ray and SZ data, using a phenomenological model, is not particularly synergistic.  All the synergy
coefficients $\zeta(CMB+XR+SZ) > 0.75$, with constraints on the dark
energy equation of state unimproved relative to adding the three
individual experiment errors in quadrature.
This implies that CMB observations resolve the degeneracies that were
previously broken by the combination of X--ray and SZ data.  
This is also the case for a physical model
with power--law $K_0$ evolution.  However, for arbitrary $K_0$
evolution, we still find significant improvement in constraints for
the combination of X--ray and SZ data, even in the presence of CMB
data, with $\zeta(CMB+XR+SZ) < 0.8$ for all parameters, and $\zeta(CMB+XR+SZ) \sim 0.7$ for dark energy equation of state parameters.  This degeneracy breaking is only slightly weaker
than in the absence of CMB data.  Indeed, though the addition of Planck data to the individual surveys yields noticeably improved constraints on dark energy, when is included in the combined surveys it
helps little with the $w_0$ and $w_a$ constraints, since their
marginalized uncertainties are dominated by the degeneracies that
remain with the cluster model parameters (rather than among
cosmological parameters).

\section{Sensitivity to the Adopted Physical Model}

While the above results are promising, one may worry that they rely on
the specific choice for the underlying physical model.  In particular,
if real data is fit with the wrong model, the derived cosmological
parameters may suffer from a bias larger than their uncertainty we
inferred above.  Such potential biases could be directly quantified by
creating mock data using various ``true" models, and obtaining the
best-fit to these data using different, parameterized, ``wrong" models.
The magnitude of the bias, in general, will depend strongly on the
pair of models that are compared -- and hence, ultimately, on which way
the real data falls.  A full exploration of these biases also requires
going beyond the Fisher matrix approach adopted here, and will be
postponed to a future paper.

Nevertheless, one can use the Fisher approach to quantify the impact
of wrong model assumptions, by adding new degrees of freedom,
representing uncertainties about the underlying physical model, and
evaluating the degradation in the statistical errors.  This
degradation can be regarded as the systematic error arising from the
choice of the wrong model, among a restriced set of possible models.

Specifically, here we allow the gas fraction to evolve as $f_g \sim
(1+z)^{\alpha_{f_g}}$, and the entropy injection to vary with mass as
$K_0 \sim M^{\kappa_M}$ and radius as $K_0 \sim (1+x)^{\kappa_x}$.  To
maintain our rough agreement with observations, the fiducial model is
chosen to match that used for the previous analysis, with
$\alpha_{f_{g}} = \kappa_M = \kappa_x = 0$, and the three additional
parameters included in our self--calibrated Fisher forecasts for a
power--law evolving entropy injection with 20 flux bins and a
log--normal scatter.

We then compute the constraints for cosmological parameters as before,
to see the extended to which they are degraded by the addition of
these new degrees of freedom. We find that in the X--ray survey, the
cosmological parameter constraints degrade by about a factor of $\sim
2$, while in the SZ survey, the constraints degrade by no more than
$15\%$.  This behavior is to be expected, as the X--ray luminosity
scales as $\sim \rho_g^2$, as compared to the SZ luminosity that scales
as $\rho_g$, and is therefore more sensitive to changes in the
physical model.  Likewise, the model parameter constraints degrade by
$\sim 60\%$ in both the X--ray and SZ surveys.  When both surveys are
combined, the cosmological parameter constraints degrade by no more
than $\sim 20\%$ -- $\lsim 10 \%$ for the dark energy equation of
state -- while the model parameter constraints degrade by $\sim 50\%$.
At the same time, we are able to constrain $\alpha_{f_g}$ -- the
evolution of the gas fraction -- to within 0.14 and 0.11 from the
X--ray and SZ surveys individually, and 0.06 from their combination;
roughly similar to constraints on the evolution of the entropy
injection.  Likewise, we are able to constrain the values of the new
parameters to within $\sim 0.02$ from the X--ray and SZ surveys
individually, and 0.01 from the combination.

Therefore, we conclude that the greatest impact of introducing
additional flexibility to the model is to degrade theoretical
constraints on the model parameters themselves.  The cosmological
constraints, particularly for the SZ and combined surveys, are more
robust to relaxing the model assumptions.  Using a model with three
extra degrees of freedom, even in the case of the X--ray survey, which
is particularly sensitive to the model behavior, the theoretical
constraints using a physical model are substantially better than those
using a phenomenological fit.  While these results are encouraging, a
more complete treatment, allowing for a wider range of
(non--power--law) models, would be necessary to estimate possible
systematic errors.

\section{Future Work}

Though these results above are promising, we here note some important
reservations.  Firstly, the goal in our study is to motivate a
methodology, rather than a specific cluster structure model.  We
believe that the improvement in the constraints in individual
experiments arises because adopting a physical model introduces new
cosmology sensitivity that is not present when the mass--observable
relation is parametrized directly. For the preheated model we
consider here, the dominant effect is the new $\Omega_m
h^2$--sensitivity, which we expect to hold generically, regardless of
the details of a cluster model.  Likewise, the improvement in the
synergy between the SZ and X--ray surveys arises because the physical
model relates the X--ray flux and the SZ decrement (one cannot be
changed without changing the other).  This, again, should be a generic
feature of any model that directly models the ICM gas distribution.  A caveat to this last conclusion is that the model may have uncertainties, such as gas clumping or a strong radial dependence of the entropy injection $K_0(r)$, which will changed the X--ray flux relative to the SZ decrement, which will tend to reduce the synergy between the two surveys.

Nevertheless, before fits to real data can be performed, there is
significant work to be done in modeling cluster structure and
evolution more accurately.  Modified entropy models, though very
successful at predicting mass--observable relations and their
evolution, are incorrect in detail \citep{pratt2005,pratt2006}.  We
also find some tension between our model predictions and observed
local scaling relations and number counts (see
Figures~\ref{fig:localscaling}, and \ref{fig:logNlogS}).
Furthermore, in addition to having models that fit cluster structure
data accurately, one has to also know the correct cosmological
dependence of the observables in these models.  In the future, we
envision that three--dimensional simulations, including the relevant
non--gravitational and gas physics, at least in parametrized manner,
and also including asymmetries in cluster structure, as well as
deviations from hydrostatic equilibrium, will provide such models.

In addition to an improved model of cluster structure, more realistic
constraints require a detailed treatment of the systematics of any
real survey.  Our work represents a ``best case scenario", in which
the X--ray and SZ surveys are idealized in many ways.  With the
introduction of scatter, and using bin sizes far wider than any
reasonable observational error on flux measurements, the inclusion of
systematic uncertainties in flux measurements will not significantly
change our overall results.  On the other hand, completeness and
selection issues will degrade final errors, and may effect the
complementarity we have found. As large-scale cluster surveys start
taking data, a more thorough understanding of the combination of the
systematic and statistical uncertainties will be required to interpret
our results.

\section{Conclusions}
\label{sec:conclude}

Using a Fisher Matrix approach, we have demonstrated that the use of a
physically motivated model of the ICM in mock large-scale cluster
surveys gives significantly better constraints on cosmological and
model parameters, and better synergy between SZ and X--ray surveys,
than one can obtain by directly parametrizing the mass--observable
relation.  In particular, when both the cosmology and model parameters
are included in the fit, for the simplest case of pure cluster counts
as a function of redshift (see Table~\ref{table:1bin0scat.cal}), the
physical model yields constraints on the dark energy equation of state
that are 2 times tighter and 3 times less degenerate in the X--ray,
and 10 times tighter and 15 times less degenerate in the SZ than those
using a phenomenological model.  If the shape of the mass function and
scatter in the mass-observable relation are included and the entropy
floor is taken to be an arbitrary function of redshift (see
Table~\ref{table:10bin10scat.cal}), the dark energy parameter
constraints are 20\% times tighter and 2 times less degenerate in the
X--ray, and 2 times tighter and 2 times less degenerate in the SZ than those
using a phenomenological model.  In addition, these constraints are up to a
factor of two tighter than those from simply adding the individual
experiment errors in quadrature, relative to a minor 20\% improvement
from combining constraints using a phenomenological model.  These result
suggest that parametrized physical models of cluster structure will
be useful when extracting constraints on both cosmology, and cluster
structure itself, from future large--scale SZ and X--ray cluster
surveys.

\acknowledgements

We thank Justin Khoury for providing the Planck Fisher matrix, and the referee for many helpful comments.  This work was supported in part by NSF grant AST-05-07161, by the
U.S. Department of Energy under Contract No. DE--AC02--98CH10886, and
by the Columbia University Initiatives in Science and Engineering
(ISE) funds.  Greg Bryan acknowledges support from NSF grant AST-0547823.

\clearpage
\newpage

\begin{table}
%tablenum{1a}
\begin{center}
\caption{Cosmological Cluster Constraints for $n_{bin} = 1$ and no scatter\label{table:1bin0scat.cosmo}}
\begin{tabular}{cccccccc}
\hline
\hline
& & $\Delta p(XR)$ & $\xi(XR)$ & $\Delta p(SZ)$ & $\xi(SZ)$ & $\Delta p(XR+SZ)$ & $\zeta(XR+SZ)$ \\ 
\hline
Phenomenological Model & $\Delta\Omega_mh^2$ & $5.01 \times 10^{-3}$ & 2.4  & $8.98 \times 10^{-3}$ & 4.9  & $2.80 \times 10^{-3}$ & 0.64  \\
 & $\Delta\Omega_{DE}$ & $7.44 \times 10^{-3}$ & 10 & $9.40 \times 10^{-3}$ & 11 & $2.43 \times 10^{-3}$ & 0.42  \\
& $\Delta w_0$ & 0.10  & 9.0  & 0.17  & 17 & $5.14 \times 10^{-2}$ & 0.59  \\
& $\Delta w_a$ & 0.31  & 5.5  & 0.57  & 11 & 0.21  & 0.78  \\
& $\Delta \sigma_8$ & $1.31 \times 10^{-2}$ & 15 & $1.31 \times 10^{-2}$ & 19 & $3.69 \times 10^{-3}$ & 0.40  \\

\hline
Physical Model & $\Delta\Omega_mh^2$ & $2.67 \times 10^{-3}$ & 1.6  & $2.10 \times 10^{-3}$ & 3.0  & $1.40 \times 10^{-3}$ & 0.85  \\
 & $\Delta\Omega_{DE}$ & $4.68 \times 10^{-3}$ & 4.0  & $4.25 \times 10^{-3}$ & 5.5  & $3.04 \times 10^{-3}$ & 0.97  \\
& $\Delta w_0$ & $4.90 \times 10^{-2}$ & 2.9  & $3.35 \times 10^{-2}$ & 3.0  & $2.64 \times 10^{-2}$ & 0.96  \\
& $\Delta w_a$ & 0.19  & 1.8  & $9.89 \times 10^{-2}$ & 1.8  & $8.34 \times 10^{-2}$ & 0.96  \\
& $\Delta \sigma_8$ & $4.33 \times 10^{-3}$ & 4.5  & $3.98 \times 10^{-3}$ & 5.6  & $2.75 \times 10^{-3}$ & 0.94  \\

\hline
\hline
\end{tabular}
\end{center}
\end{table}

\begin{table}
%tablenum{1b}
\begin{center}
\caption{Cosmological Cluster Constraints for $n_{bin} = 20$ and no scatter\label{table:10bin0scat.cosmo}}
\begin{tabular}{cccccccc}
\hline
\hline
& & $\Delta p(XR)$ & $\xi(XR)$ & $\Delta p(SZ)$ & $\xi(SZ)$ & $\Delta p(XR+SZ)$ & $\zeta(XR+SZ)$ \\ 
\hline
Phenomenological Model & $\Delta\Omega_mh^2$ & $7.08 \times 10^{-4}$ & 1.1  & $1.22 \times 10^{-3}$ & 1.3  & $6.06 \times 10^{-4}$ & 0.99  \\
 & $\Delta\Omega_{DE}$ & $8.07 \times 10^{-4}$ & 1.5  & $1.15 \times 10^{-3}$ & 1.8  & $6.32 \times 10^{-4}$ & 0.96  \\
& $\Delta w_0$ & $6.85 \times 10^{-3}$ & 1.2  & $7.27 \times 10^{-3}$ & 1.3  & $4.89 \times 10^{-3}$ & 0.98  \\
& $\Delta w_a$ & $6.45 \times 10^{-3}$ & 1.0  & $9.05 \times 10^{-3}$ & 1.2  & $5.07 \times 10^{-3}$ & 0.97  \\
& $\Delta \sigma_8$ & $1.31 \times 10^{-3}$ & 1.6  & $1.28 \times 10^{-3}$ & 1.9  & $8.49 \times 10^{-4}$ & 0.93  \\

\hline
Physical Model & $\Delta\Omega_mh^2$ & $1.44 \times 10^{-4}$ & 1.0  & $2.01 \times 10^{-4}$ & 1.1  & $1.17 \times 10^{-4}$ & 0.99  \\
 & $\Delta\Omega_{DE}$ & $7.10 \times 10^{-4}$ & 1.1  & $6.15 \times 10^{-4}$ & 1.3  & $4.53 \times 10^{-4}$ & 0.98  \\
& $\Delta w_0$ & $9.21 \times 10^{-3}$ & 1.1  & $1.01 \times 10^{-2}$ & 1.2  & $6.61 \times 10^{-3}$ & 0.97  \\
& $\Delta w_a$ & $1.22 \times 10^{-2}$ & 1.1  & $9.55 \times 10^{-3}$ & 1.1  & $7.20 \times 10^{-3}$ & 0.96  \\
& $\Delta \sigma_8$ & $1.06 \times 10^{-3}$ & 1.2  & $1.06 \times 10^{-3}$ & 1.5  & $7.01 \times 10^{-4}$ & 0.93  \\

\hline
\hline
\end{tabular}
\end{center}
\end{table}
\newpage
\begin{table}
%tablenum{1c}
\begin{center}
\caption{Cosmological Cluster Constraints for $n_{bin} = 20$ and $\sigma_{\log{M}|\Psi} = 0.1$ log-Normal scatter\label{table:10bin10scat.cosmo}}
\begin{tabular}{cccccccc}
\hline
\hline
& & $\Delta p(XR)$ & $\xi(XR)$ & $\Delta p(SZ)$ & $\xi(SZ)$ & $\Delta p(XR+SZ)$ & $\zeta(XR+SZ)$ \\ 
\hline
Phenomenological & $\Delta\Omega_mh^2$ & $1.94 \times 10^{-3}$ & 2.3  & $5.08 \times 10^{-3}$ & 3.7  & $1.31 \times 10^{-3}$ & 0.72  \\
 & $\Delta\Omega_{DE}$ & $4.23 \times 10^{-3}$ & 6.5  & $7.21 \times 10^{-3}$ & 9.1  & $2.07 \times 10^{-3}$ & 0.57  \\
& $\Delta w_0$ & $6.21 \times 10^{-2}$ & 6.0  & $9.15 \times 10^{-2}$ & 10 & $3.94 \times 10^{-2}$ & 0.77  \\
& $\Delta w_a$ & 0.23  & 4.2  & 0.32  & 6.7  & 0.16  & 0.86  \\
& $\Delta \sigma_8$ & $7.26 \times 10^{-3}$ & 9.0  & $8.57 \times 10^{-3}$ & 13 & $2.83 \times 10^{-3}$ & 0.51  \\
& $\Delta\sigma_{XR}$ & $1.22 \times 10^{-3}$ & \nodata & \nodata & \nodata & $1.20 \times 10^{-3}$ & \nodata \\
& $\Delta\sigma_{SZ}$ & \nodata & \nodata & $5.73 \times 10^{-3}$ & \nodata & $3.11 \times 10^{-3}$ & \nodata \\

\hline
Physical Model & $\Delta\Omega_mh^2$ & $1.55 \times 10^{-3}$ & 1.6  & $1.51 \times 10^{-3}$ & 2.3  & $9.99 \times 10^{-4}$ & 0.92  \\
 & $\Delta\Omega_{DE}$ & $2.75 \times 10^{-3}$ & 2.7  & $3.81 \times 10^{-3}$ & 5.1  & $2.11 \times 10^{-3}$ & 0.95  \\
& $\Delta w_0$ & $3.62 \times 10^{-2}$ & 2.5  & $3.01 \times 10^{-2}$ & 3.0  & $2.02 \times 10^{-2}$ & 0.87  \\
& $\Delta w_a$ & 0.12  & 1.5  & $9.38 \times 10^{-2}$ & 1.8  & $7.16 \times 10^{-2}$ & 0.96  \\
& $\Delta \sigma_8$ & $8.01 \times 10^{-3}$ & 8.7  & $4.19 \times 10^{-3}$ & 6.1  & $2.39 \times 10^{-3}$ & 0.64  \\
& $\Delta \sigma_{XR}$ & $1.61 \times 10^{-2}$ & \nodata & \nodata & \nodata & $9.90 \times 10^{-3}$ & \nodata \\
& $\Delta \sigma_{SZ}$ & \nodata & \nodata & $2.45 \times 10^{-2}$ & \nodata & 1.4  & \nodata \\

\hline
\hline
\end{tabular}
\end{center}
\end{table}

\begin{table}
%tablenum{2a}
\begin{center}
\caption{Self-Calibrated Cluster Constraints for $n_{bin} = 1$ with no scatter\label{table:1bin0scat.cal}
.}
\begin{tabular}{cccccccc}
\hline
\hline
& & $\Delta p(XR)$ & $\xi(XR)$ & $\Delta p(SZ)$ & $\xi(SZ)$ & $\Delta p(XR+SZ)$ & $\zeta(XR+SZ)$ \\ 
\hline
Phenomenological Model & $\Delta\Omega_mh^2$ & $2.73 \times 10^{-2}$ & 13 & 0.15  & 82 & $1.18 \times 10^{-2}$ & 0.44  \\
 & $\Delta\Omega_{DE}$ & $5.97 \times 10^{-2}$ & 85 & 0.17  & 206 & $2.19 \times 10^{-2}$ & 0.39  \\
& $\Delta w_0$ & 0.21  & 18 & 0.50  & 50 & 0.15  & 0.80  \\
& $\Delta w_a$ & 0.32  & 5.7  & 1.1  & 21 & 0.27  & 0.88  \\
& $\Delta \sigma_8$ & $8.31 \times 10^{-2}$ & 98 & 0.43  & 627 & $5.51 \times 10^{-2}$ & 0.67  \\
& $\Delta\log{A_x}$ & 0.26  & \nodata & \nodata & \nodata & 0.23  & \nodata \\
& $\Delta\beta_x$ & 0.21  & \nodata & \nodata & \nodata & 0.15  & \nodata \\
& $\Delta\gamma_x$ & 0.64  & \nodata & \nodata & \nodata & 0.41  & \nodata \\
& $\Delta\log{A_{sz}}$ & \nodata & \nodata & 1.3  & \nodata & 0.14  & \nodata \\
& $\Delta\beta_{sz}$ & \nodata & \nodata & 0.83  & \nodata & 0.10  & \nodata \\
& $\Delta\gamma_{sz}$ & \nodata & \nodata & 2.1  & \nodata & 0.53  & \nodata \\

\hline
Physical Model & $\Delta\Omega_mh^2$ & $4.01 \times 10^{-3}$ & 2.3  & $2.18 \times 10^{-3}$ & 3.1  & $1.66 \times 10^{-3}$ & 0.87  \\
$K_0(z) = K_0(1+z)^{\alpha_K}$ & $\Delta\Omega_{DE}$ & $1.37 \times 10^{-2}$ & 11 & $4.65 \times 10^{-3}$ & 6.1  & $3.63 \times 10^{-3}$ & 0.82  \\
& $\Delta w_0$ & 0.12  & 7.0  & $3.39 \times 10^{-2}$ & 3.1  & $2.84 \times 10^{-2}$ & 0.87  \\
& $\Delta w_a$ & 0.19  & 1.8  & 0.10  & 1.8  & $8.54 \times 10^{-2}$ & 0.96  \\
& $\Delta \sigma_8$ & $2.01 \times 10^{-2}$ & 20 & $5.77 \times 10^{-3}$ & 8.1  & $4.28 \times 10^{-3}$ & 0.77  \\
& $\Delta f_g$ & $7.07 \times 10^{-2}$ & \nodata & $3.76 \times 10^{-2}$ & \nodata & $2.25 \times 10^{-2}$ & \nodata \\
& $\Delta K_0$ & 18 & \nodata & 6.2  & \nodata & 3.0  & \nodata \\
& $\Delta \alpha_K$ & 0.36  & \nodata & 0.10  & \nodata & $6.03 \times 10^{-2}$ & \nodata \\
\hline 

\hline 
\hline
\end{tabular}
\end{center}
\end{table}

\begin{table}
%tablenum{2b}
\begin{center}
\caption{Self-Calibrated Cluster Constraints for $n_{bin} = 20$ with no scatter.\label{table:10bin0scat.cal}}
\begin{tabular}{cccccccc}
\hline
\hline
& & $\Delta p(XR)$ & $\xi(XR)$ & $\Delta p(SZ)$ & $\xi(SZ)$ & $\Delta p(XR+SZ)$ & $\zeta(XR+SZ)$ \\ 
\hline
Phenomenological Model & $\Delta\Omega_mh^2$ & $7.36 \times 10^{-4}$ & 1.2  & $1.78 \times 10^{-3}$ & 1.9  & $6.53 \times 10^{-4}$ & 0.96  \\

 & $\Delta\Omega_{DE}$ & $8.64 \times 10^{-4}$ & 1.6  & $1.22 \times 10^{-3}$ & 1.9  & $6.74 \times 10^{-4}$ & 0.96  \\
& $\Delta w_0$ & $7.19 \times 10^{-3}$ & 1.3  & $8.03 \times 10^{-3}$ & 1.4  & $5.28 \times 10^{-3}$ & 0.99  \\
& $\Delta w_a$ & $6.53 \times 10^{-3}$ & 1.0  & $9.31 \times 10^{-3}$ & 1.2  & $5.16 \times 10^{-3}$ & 0.96  \\
& $\Delta \sigma_8$ & $1.76 \times 10^{-3}$ & 2.2  & $2.13 \times 10^{-3}$ & 3.2  & $1.30 \times 10^{-3}$ & 0.96  \\
& $\Delta\log{A_x}$ & $1.89 \times 10^{-3}$ & \nodata & \nodata & \nodata & $1.74 \times 10^{-3}$ & \nodata \\
& $\Delta\beta_x$ & $3.42 \times 10^{-3}$ & \nodata & \nodata & \nodata & $2.95 \times 10^{-3}$ & \nodata \\
& $\Delta\gamma_x$ & $9.23 \times 10^{-4}$ & \nodata & \nodata & \nodata & $9.01 \times 10^{-4}$ & \nodata \\
& $\Delta\log{A_{sz}}$ & \nodata & \nodata & $4.08 \times 10^{-3}$ & \nodata & $3.20 \times 10^{-3}$ & \nodata \\
& $\Delta\beta_{sz}$ & \nodata & \nodata & $5.69 \times 10^{-3}$ & \nodata & $4.39 \times 10^{-3}$ & \nodata \\
& $\Delta\gamma_{sz}$ & \nodata & \nodata & $5.07 \times 10^{-4}$ & \nodata & $4.81 \times 10^{-4}$ & \nodata \\

\hline
Physical Model & $\Delta\Omega_mh^2$ & $1.48 \times 10^{-4}$ & 1.1  & $2.08 \times 10^{-4}$ & 1.1  & $1.17 \times 10^{-4}$ & 0.98  \\
$K_0(z) = K_0(1+z)^{\alpha_K}$ & $\Delta\Omega_{DE}$ & $7.40 \times 10^{-4}$ & 1.1  & $6.16 \times 10^{-4}$ & 1.3  & $4.59 \times 10^{-4}$ & 0.97  \\
& $\Delta w_0$ & $1.00 \times 10^{-2}$ & 1.2  & $1.05 \times 10^{-2}$ & 1.3  & $6.87 \times 10^{-3}$ & 0.95  \\
& $\Delta w_a$ & $1.23 \times 10^{-2}$ & 1.1  & $1.05 \times 10^{-2}$ & 1.2  & $7.46 \times 10^{-3}$ & 0.93  \\
& $\Delta \sigma_8$ & $1.88 \times 10^{-3}$ & 2.0  & $1.21 \times 10^{-3}$ & 1.7  & $8.63 \times 10^{-4}$ & 0.85  \\
& $\Delta f_g$ & $4.18 \times 10^{-3}$ & \nodata & $3.60 \times 10^{-3}$ & \nodata & $2.48 \times 10^{-3}$ & \nodata \\
& $\Delta K_0$ & 0.74  & \nodata & 0.91  & \nodata & 0.52  & \nodata \\
& $\Delta\alpha_K$ & $2.19 \times 10^{-2}$ & \nodata & $1.64 \times 10^{-2}$ & \nodata & $1.15 \times 10^{-2}$ & \nodata \\
\hline
Physical Model & $\Delta\Omega_mh^2$ & $1.95 \times 10^{-4}$ & 1.4  & $2.53 \times 10^{-4}$ & 1.3  & $1.32 \times 10^{-4}$ & 0.86  \\
Arbitrary $K_0(z)$ & $\Delta\Omega_{DE}$ & $8.34 \times 10^{-4}$ & 1.3  & $8.69 \times 10^{-4}$ & 1.9  & $4.90 \times 10^{-4}$ & 0.81  \\
& $\Delta w_0$ & $1.15 \times 10^{-2}$ & 1.4  & $1.37 \times 10^{-2}$ & 1.7  & $8.08 \times 10^{-3}$ & 0.91  \\
& $\Delta w_a$ & $1.62 \times 10^{-2}$ & 1.4  & $1.48 \times 10^{-2}$ & 1.6  & $8.68 \times 10^{-3}$ & 0.79  \\
& $\Delta \sigma_8$ & $2.08 \times 10^{-3}$ & 2.3  & $1.62 \times 10^{-3}$ & 2.3  & $9.82 \times 10^{-4}$ & 0.77  \\
& $\Delta f_g$ & $4.48 \times 10^{-3}$ & \nodata & $4.10 \times 10^{-3}$ & \nodata & $2.70 \times 10^{-3}$ & \nodata \\
& $<\Delta K_0>$ & 1.3  & \nodata & 1.2  & \nodata & 0.68  & \nodata \\

\hline
\hline
\end{tabular}
\end{center}
\end{table}

\begin{table}
%tablenum{2c}
\begin{center}
\caption{Self-Calibrated Cluster Constraints for $n_{bin} = 20$ and $\sigma_{\log{M}|\Psi} = 0.1$ log-Normal scatter.\label{table:10bin10scat.cal}}
\begin{tabular}{cccccccc}
\hline
\hline
& & $\Delta p(XR)$ & $\xi(XR)$ & $\Delta p(SZ)$ & $\xi(SZ)$ & $\Delta p(XR+SZ)$ & $\zeta(XR+SZ)$ \\ 
\hline
Phenomenological & $\Delta\Omega_mh^2$ & $5.93 \times 10^{-3}$ & 6.9  & $2.89 \times 10^{-2}$ & 21 & $2.58 \times 10^{-3}$ & 0.44  \\
 & $\Delta\Omega_{DE}$ & $1.43 \times 10^{-2}$ & 22 & $1.32 \times 10^{-2}$ & 16 & $5.66 \times 10^{-3}$ & 0.58  \\
& $\Delta w_0$ & $7.88 \times 10^{-2}$ & 7.7  & 0.11  & 11 & $5.55 \times 10^{-2}$ & 0.88  \\
& $\Delta w_a$ & 0.27  & 4.9  & 0.34  & 7.1  & 0.19  & 0.91  \\
& $\Delta \sigma_8$ & $1.40 \times 10^{-2}$ & 17 & $1.56 \times 10^{-2}$ & 23 & $8.85 \times 10^{-3}$ & 0.85  \\
& $\Delta\log{A_x}$ & $2.56 \times 10^{-2}$ & \nodata & \nodata & \nodata & $1.72 \times 10^{-2}$ & \nodata \\
& $\Delta\beta_x$ & $2.09 \times 10^{-2}$ & \nodata & \nodata & \nodata & $1.22 \times 10^{-2}$ & \nodata \\
& $\Delta\gamma_x$ & $9.67 \times 10^{-2}$ & \nodata & \nodata & \nodata & $5.80 \times 10^{-2}$ & \nodata \\
& $\Delta\sigma_{XR}$ & $1.24 \times 10^{-3}$ & \nodata & \nodata & \nodata & $1.23 \times 10^{-3}$ & \nodata \\
& $\Delta\log{A_{sz}}$ & \nodata & \nodata & $3.20 \times 10^{-2}$ & \nodata & $2.12 \times 10^{-2}$ & \nodata \\
& $\Delta\beta_{sz}$ & \nodata & \nodata & $4.77 \times 10^{-2}$ & \nodata & $1.36 \times 10^{-2}$ & \nodata \\
& $\Delta\gamma_{sz}$ & \nodata & \nodata & $3.32 \times 10^{-2}$ & \nodata & $2.80 \times 10^{-2}$ & \nodata \\
& $\Delta\sigma_{SZ}$ & \nodata & \nodata & $6.25 \times 10^{-3}$ & \nodata & $5.71 \times 10^{-3}$ & \nodata \\

\hline
Physical Model & $\Delta\Omega_mh^2$ & $1.66 \times 10^{-3}$ & 1.7  & $1.55 \times 10^{-3}$ & 2.4  & $1.05 \times 10^{-3}$ & 0.92  \\
$K_0(z) = K_0(1+z)^{\alpha_K}$ & $\Delta\Omega_{DE}$ & $2.86 \times 10^{-3}$ & 2.8  & $4.11 \times 10^{-3}$ & 5.5  & $2.22 \times 10^{-3}$ & 0.95  \\
& $\Delta w_0$ & $5.18 \times 10^{-2}$ & 3.5  & $3.03 \times 10^{-2}$ & 3.0  & $2.16 \times 10^{-2}$ & 0.83  \\
& $\Delta w_a$ & 0.13  & 1.5  & $9.60 \times 10^{-2}$ & 1.9  & $7.30 \times 10^{-2}$ & 0.95  \\
& $\Delta \sigma_8$ & $9.40 \times 10^{-3}$ & 10 & $5.31 \times 10^{-3}$ & 7.8  & $3.33 \times 10^{-3}$ & 0.72  \\
& $\Delta \sigma_{XR}$ & $6.47 \times 10^{-3}$ & \nodata & \nodata & \nodata & $4.89 \times 10^{-3}$ & \nodata \\
& $\Delta \sigma_{SZ}$ & \nodata & \nodata & $7.55 \times 10^{-3}$ & \nodata & $6.30 \times 10^{-3}$ & \nodata \\
& $\Delta f_g$ & $2.09 \times 10^{-2}$ & \nodata & $2.67 \times 10^{-2}$ & \nodata & $1.10 \times 10^{-2}$ & \nodata \\
& $\Delta K_0$ & 5.0  & \nodata & 3.8  & \nodata & 2.2  & \nodata \\
& $\Delta\alpha_K$ & $8.79 \times 10^{-2}$ & \nodata & $7.65 \times 10^{-2}$ & \nodata & $4.37 \times 10^{-2}$ & \nodata \\

\hline
Physical Model & $\Delta\Omega_mh^2$ & $2.12 \times 10^{-3}$ & 2.1  & $2.17 \times 10^{-3}$ & 3.3  & $1.19 \times 10^{-3}$ & 0.79  \\
Arbitrary $K_0(z)$ & $\Delta\Omega_{DE}$ & $3.66 \times 10^{-3}$ & 3.6  & $7.11 \times 10^{-3}$ & 9.6  & $2.56 \times 10^{-3}$ & 0.79  \\
& $\Delta w_0$ & $7.73 \times 10^{-2}$ & 5.2  & $6.05 \times 10^{-2}$ & 5.9  & $2.61 \times 10^{-2}$ & 0.55  \\
& $\Delta w_a$ & 0.23  & 2.7  & 0.17  & 3.3  & $9.25 \times 10^{-2}$ & 0.68  \\
& $\Delta \sigma_8$ & $1.45 \times 10^{-2}$ & 15 & $7.85 \times 10^{-3}$ & 11 & $3.62 \times 10^{-3}$ & 0.52  \\
& $\Delta \sigma_{XR}$ & $7.79 \times 10^{-3}$ & \nodata & \nodata & \nodata & $5.61 \times 10^{-3}$ & \nodata \\
& $\Delta \sigma_{SZ}$ & \nodata & \nodata & $9.77 \times 10^{-3}$ & \nodata & $6.74 \times 10^{-3}$ & \nodata \\
& $\Delta f_g$ & $2.48 \times 10^{-2}$ & \nodata & $3.94 \times 10^{-2}$ & \nodata & $1.23 \times 10^{-2}$ & \nodata \\
& $<\Delta K_0>$ & 2.2  & \nodata & 2.8  & \nodata & 1.5  & \nodata \\

\hline 
\hline
\end{tabular}
\end{center}
\end{table}

\begin{table}
%tablenum{3}
\begin{center}
\caption{Cosmological Constraints for CMB Observations\label{table:complementary}.}
\begin{tabular}{ccccc}
\hline
\hline
&  $\Delta p(PLANCK)$ & $\xi(PLANCK)$ \\ 
\hline
$\Delta \Omega_mh^2$ & $1.21 \times 10^{-3}$ & 23 \\
$\Delta \Omega_{DE}$ & $3.46 \times 10^{-2}$ & 72 \\
$\Delta w_0$ & 0.32 & 166 \\
$\Delta w_a$ & 1.04 & 152  \\
$\Delta \sigma_8$ & $4.11 \times 10^{-2}$ & 88  \\
$\Delta \Omega_b h^2$ & $1.36\times 10^{-4}$ & \nodata \\
$\Delta n_s$ & $3.54\times 10^{-3}$ & \nodata  \\
\hline 
\hline
\end{tabular}
\end{center}
\end{table}

\begin{table}
%tablenum{4}
\begin{center}
\caption{Self-Calibrated Cluster+CMB Constraints for $n_{bin} = 20$ and $\sigma_{\log{M}|\Psi} = 0.1$ log-Normal scatter.\label{table:10bin10scat.cmb}}
\begin{tabular}{cccccccc}
\hline
\hline
& & $\Delta p(XR)$ & $\xi(XR)$ & $\Delta p(SZ)$ & $\xi(SZ)$ & $\Delta p(XR+SZ)$ & $\zeta(XR+SZ)$ \\ 
\hline
Power-Law Model & $\Delta\Omega_mh^2$ & $5.93 \times 10^{-4}$ & 11 & $5.34 \times 10^{-4}$ & 10 & $3.35 \times 10^{-4}$ & 0.89  \\
 & $\Delta\Omega_{DE}$ & $3.17 \times 10^{-3}$ & 8.2  & $5.53 \times 10^{-3}$ & 13 & $2.07 \times 10^{-3}$ & 0.76  \\
& $\Delta w_0$ & $6.07 \times 10^{-2}$ & 32 & $8.67 \times 10^{-2}$ & 46 & $4.77 \times 10^{-2}$ & 0.97  \\
& $\Delta w_a$ & 0.20  & 30 & 0.25  & 37 & 0.16  & 1.0  \\
& $\Delta \sigma_8$ & $4.87 \times 10^{-3}$ & 12 & $1.12 \times 10^{-2}$ & 29 & $3.78 \times 10^{-3}$ & 0.85  \\
& $\Delta \Omega_b h^2$ & $1.10 \times 10^{-4}$ & \nodata & $1.11 \times 10^{-4}$ & \nodata & $1.07 \times 10^{-4}$ & \nodata \\
& $\Delta n_s$ & $2.56 \times 10^{-3}$ & \nodata & $2.49 \times 10^{-3}$ & \nodata & $2.30 \times 10^{-3}$ & \nodata \\
& $\Delta\log{A_x}$ & $2.21 \times 10^{-2}$ & \nodata & \nodata & \nodata & $1.07 \times 10^{-4}$ & \nodata \\
& $\Delta\beta_x$ & $9.97 \times 10^{-3}$ & \nodata & \nodata & \nodata & $2.30 \times 10^{-3}$ & \nodata \\
& $\Delta\gamma_x$ & $4.26 \times 10^{-2}$ & \nodata & \nodata & \nodata & $1.67 \times 10^{-2}$ & \nodata \\
& $\Delta\sigma_{XR}$ & $1.22 \times 10^{-3}$ & \nodata & \nodata & \nodata & $1.22 \times 10^{-3}$ & \nodata \\
& $\Delta\log{A_{sz}}$ & \nodata & \nodata & $2.71 \times 10^{-2}$ & \nodata & $1.82 \times 10^{-2}$ & \nodata \\
& $\Delta\beta_{sz}$ & \nodata & \nodata & $1.32 \times 10^{-2}$ & \nodata & $1.22 \times 10^{-2}$ & \nodata \\
& $\Delta\gamma_{sz}$ & \nodata & \nodata & $2.50 \times 10^{-2}$ & \nodata & $2.28 \times 10^{-2}$ & \nodata \\
& $\Delta\sigma_{SZ}$ & \nodata & \nodata & $5.00 \times 10^{-3}$ & \nodata & $4.50 \times 10^{-3}$ & \nodata \\

\hline
Physical Model & $\Delta\Omega_mh^2$ & $5.72 \times 10^{-4}$ & 10 & $3.79 \times 10^{-4}$ & 7.2  & $2.97 \times 10^{-4}$ & 0.97  \\
$K_0(z) = K_0(1+z)^{\alpha_K}$ & $\Delta\Omega_{DE}$ & $2.39 \times 10^{-3}$ & 5.5  & $3.61 \times 10^{-3}$ & 9.0  & $1.85 \times 10^{-3}$ & 0.93  \\
& $\Delta w_0$ & $3.18 \times 10^{-2}$ & 16 & $2.85 \times 10^{-2}$ & 15 & $1.98 \times 10^{-2}$ & 0.93  \\
& $\Delta w_a$ & 0.12  & 17 & $8.44 \times 10^{-2}$ & 12 & $6.67 \times 10^{-2}$ & 0.97  \\
& $\Delta \sigma_8$ & $5.91 \times 10^{-3}$ & 14 & $5.05 \times 10^{-3}$ & 13 & $3.03 \times 10^{-3}$ & 0.79  \\
& $\Delta \Omega_b h^2$ & $1.13 \times 10^{-4}$ & \nodata & $1.05 \times 10^{-4}$ & \nodata & $1.03 \times 10^{-4}$ & \nodata \\
& $\Delta n_s$ & $2.54 \times 10^{-3}$ & \nodata & $2.33 \times 10^{-3}$ & \nodata & $2.26 \times 10^{-3}$ & \nodata \\
& $\Delta \sigma_{XR}$ & $5.59 \times 10^{-3}$ & \nodata & \nodata & \nodata & $4.74 \times 10^{-3}$ & \nodata \\
& $\Delta \sigma_{SZ}$ & \nodata & \nodata & $6.39 \times 10^{-3}$ & \nodata & $5.26 \times 10^{-3}$ & \nodata \\
& $\Delta f_g$ & $1.37 \times 10^{-2}$ & \nodata & $2.59 \times 10^{-2}$ & \nodata & $1.04 \times 10^{-2}$ & \nodata \\
& $\Delta K_0$ & 3.4  & \nodata & 3.7  & \nodata & 2.1  & \nodata \\
& $\Delta\alpha_K$ & $5.02 \times 10^{-2}$ & \nodata & $7.50 \times 10^{-2}$ & \nodata & $3.88 \times 10^{-2}$ & \nodata \\
\hline
Physical Model & $\Delta\Omega_mh^2$ & $6.70 \times 10^{-4}$ & 12 & $4.96 \times 10^{-4}$ & 9.4  & $3.04 \times 10^{-4}$ & 0.80  \\
$K_0(z) = K_0(1+z)^{\alpha_K}$ & $\Delta\Omega_{DE}$ & $2.73 \times 10^{-3}$ & 6.3  & $6.46 \times 10^{-3}$ & 16 & $2.08 \times 10^{-3}$ & 0.83  \\
& $\Delta w_0$ & $4.45 \times 10^{-2}$ & 23 & $5.53 \times 10^{-2}$ & 29 & $2.38 \times 10^{-2}$ & 0.69  \\
& $\Delta w_a$ & 0.18  & 25 & 0.15  & 22 & $7.98 \times 10^{-2}$ & 0.70  \\
& $\Delta \sigma_8$ & $7.49 \times 10^{-3}$ & 18 & $7.44 \times 10^{-3}$ & 19 & $3.37 \times 10^{-3}$ & 0.64  \\
& $\Delta \Omega_b h^2$ & $1.16 \times 10^{-4}$ & \nodata & $1.08 \times 10^{-4}$ & \nodata & $1.04 \times 10^{-4}$ & \nodata \\
& $\Delta n_s$ & $2.67 \times 10^{-3}$ & \nodata & $2.45 \times 10^{-3}$ & \nodata & $2.27 \times 10^{-3}$ & \nodata \\
& $\Delta \sigma_{XR}$ & 5.3  & \nodata & \nodata & \nodata & 6.8  & \nodata \\
& $\Delta \sigma_{SZ}$ & \nodata & \nodata & 4.1  & \nodata & 2.8  & \nodata \\
& $\Delta f_g$ & $7.03 \times 10^{-3}$ & \nodata & $8.33 \times 10^{-3}$ & \nodata & $5.51 \times 10^{-3}$ & \nodata \\
& $\Delta K_0$ & $1.62 \times 10^{-2}$ & \nodata & $3.52 \times 10^{-2}$ & \nodata & $5.81 \times 10^{-3}$ & \nodata \\
& $\Delta\alpha_K$ & 13 & \nodata & 8.3  & \nodata & $1.12 \times 10^{-2}$ & \nodata \\

\hline 
\hline
\end{tabular}
\end{center}
\end{table}


\begin{thebibliography}{}

\bibitem[Arnaud \& Evrard(1999)]{arnaud1999} Arnaud, M. \& Evrard, A. E., 1999, MNRAS, 305, 631
\bibitem[Bahcall \& Bode(2003)]{bahcallbode2003} Bahcall, N. A. \& Bode, P., 2003, ApJ, 588, L1
\bibitem[Bahcall \& Cen(1992)]{bahcall1992}Bahcall, N. A. \& Cen, R., 1992, ApJ, 398, L81
\bibitem[Bahcall et al.(2003)]{bahcall2003} Bahcall, N., et al., 2003, ApJ, 585, 182
\bibitem[Bahcall(1988)]{bahcall1988} Bahcall, N. A., 1988, ARA\&A, 26, 63
\bibitem[Blake \& Glazebrook(2003)]{blake2003} Blake, C. \& Glazebrook, K., 2003, ApJ, 594, 665
\bibitem[Borgani et al.(2001)]{borgani2001} Borgani, S., et al., 2001, ApJ, 561, 13
\bibitem[Brown et al.(2002)]{brown2002}Brown, M. L., Taylor, A. N., Bacon, D. J., Gray, M. E., Dye, S., Meisenheimer, K., \& Wolf, C., 2003, MNRAS, 341, 100
\bibitem[Bullock et al.(2001)]{bullock2001} Bullock, J. S., Kolatt, T. S., Sigad, Y., Somerville, R. S., Kravtsov, A. V., Klypin, A. A., Primack, J. R., \& Dekel, A.,  2001, MNRAS, 321, 559
\bibitem[Chevallier \& Polarski(2001)]{chevallier2001} Chevallier, M. \& Polarski, D., 2001, Int. J. Mod. Phys. D10, 213
\bibitem[Carlberg et al.(1997)]{carlberg1997} Carlberg, R. G., Morris, S. L., Yee, H. K. C., \& Ellingson, E., 1997, ApJ, 479, L19
\bibitem[Corasaniti et al.(2004)]{corasaniti2004} Corasaniti, P. S., Kunz, M., Parkinson, D., Copeland, E. J., \& Bassett, B. A., 2004, Phys. Rev. D., 70, 083006
\bibitem[Eisenstein \& Hu(1999)]{eisenstein1999} Eisenstein, D. J. \& Hu, W., 1999, ApJ, 511, 5
\bibitem[Eke, Navarro, \& Steinmetz(2001)]{eke2001} Eke, V. R., Navarro, J. F., \& Steinmetz, M., 2001, ApJ, 554, 114
\bibitem[Fairley et al.(2000)]{fairley2000} Fairley, B. W., Jones, L. R., Scharf, C., Ebeling, H., Perlman, E., Horner, D., Wegner, G., \& Malkan, M., 2000, MNRAS, 315, 669
\bibitem[Finoguenov, Reiprich \& B\"{o}hringer(2001)]{finoguenov01} Finoguenov, A., Reiprich, T. H., \& B\"{o}hringer, H. 2001, A\&A, 368, 749
\bibitem[Fixsen et al.(1996)]{fixsen1996} Fixsen, D. J., et al., 1996, ApJ, 473, 576
\bibitem[Freese, Adams, \& Frieman(1987)]{freese1987} Freese, K., Adams, F. C., \& Frieman, J. A., 1987, Nucl. Phys. B, 298, 797
\bibitem[Giavalisco et al.(2004)]{giavalisco2004} Giavalisco, M., et al., 2004, ApJ, 600, L103
\bibitem[Gioia  et al.(2001)]{gioia2001} Gioia, I. M., Henry, J. P., Mullis, C. R., Voges, W., Briel, U. G., B\"{o}hringer, H., \& Huchra, J. P., 2001, ApJ, 533, L105
\bibitem[Girardi et al.(1998)]{girardi1998} Girardi, M., Giuricin, G., Mardirossian, F., Mezzetti, M., \& Boschin, W., 1998, ApJ, 505, 74
\bibitem[Gladders et al.(2006)]{gladders2006} Gladders, M. D., Yee, H. K. C., Majumdar, S., Barrientos, L. F., Hoekstra, H., Hall, P. B., \& Infante, L., astro-ph/0603588
\bibitem[Haiman et al.(2005)]{haiman2005} Haiman, Z, et al., 2005, astro-ph/0507013
\bibitem[Haiman, Mohr, \& Holder(2001)]{haiman2001} Haiman, Z., Mohr, J. J., \& Holder, G. P., 2001, ApJ, 553, 545
\bibitem[Helsdon \& Ponman(2000)]{helsdon2000} Helsdon, S. F. \& Ponman, T. J., 2000, MNRAS, 319, 933
\bibitem[Henry \& Arnaud(1991)]{henry1991} Henry, J. P. \& Arnaud, K. A., 1991, ApJ, 372, 410
\bibitem[Henry(2004)]{henry2004} Henry, J. P., 2004, ApJ, 609, 603
\bibitem[Hoekstra et al.(2002)]{hoekstra2002} Hoekstra, H., van Waerbeke, L., Gladders, M. D., Mellier, Y. \& Yee, H. K. C., 2002, ApJ, 577, 604
\bibitem[Holder, Haiman, \& Mohr(2001)]{holder2001} Holder, G. P., Haiman, Z., \&  Mohr, J. J., 2001, ApJ, 560, L111
\bibitem[Horner, Mushotzky \& Scharf(1999)]{horner99} Horner, D. J., Mushotzky, R. F. \& Scharf, C. A., 1999, ApJ, 520, 78
\bibitem[Hu \& Haiman(2003)]{huhaiman2003} Hu, W. \& Haiman, Z., 2003, Phys. Rev. D, 68, 3004
\bibitem[Hu \& Kravtsov(2003)]{hukrav2003} Hu, W. \& Kravtsov, A. V., 2003, ApJ, 584, 702
\bibitem[Hu(2003)]{hu2003} Hu, W., 2003, Phys. Rev. D, 67, 081304
\bibitem[Jahoda(2003)]{jahoda2003} see Jahoda, K., 2003, Astronomische Nachrichten, 324, 132
\bibitem[Jarvis et al.(2003)]{jarvis2003} Jarvis, M., Bernstein, G. M., Fischer, P., Jain, B., Tyson, J. A., \& Wittman, D., 2003, AJ, 125, 1014
\bibitem[Jenkins et al.(2001)]{jenkins2001} Jenkins, A., Frenk, C. S., White, S. D. M., Colberg, J. M., Cole, S.,  Evrard, A. E., Couchman, H. M. P., \& Yoshida, N., 2001, MNRAS, 321, 372
\bibitem[Kuhlen et al.(2005)]{kuhlen2005} Kuhlen, M., Strigari, L., Zetnet, A., Bullock, J., \& Primack, J., 2005, MNRAS, 357, 387
\bibitem[Levine, Schultz, \& White(2002)]{levine2002} Levine, E. S., Schultz, A. E., \& White, M., 2002, ApJ, 577, 569
\bibitem[Lima \& Hu(2005)]{lima2005} Lima, M. \& Hu, W., 2005, Phys. Rev. D., 72, 043006
\bibitem[Linder(2003)]{linder2003} Linder, E. V., 2003, Phys. Rev. D., 68, 3504
\bibitem[Lloyd-Davies, Ponman, \& Cannon(2000)]{lloyddavies2000} Lloyd-Davies, E. J., Ponman, T. J., \& Cannon, D. B., 2000, MNRAS, 315, 689
\bibitem[Majumdar \& Mohr(2004)]{majumdar2004} Majumdar, S. \& Mohr, J. J., 2004, ApJ, 613, 41
\bibitem[Markevich(1998)]{markevich1998} Markevich, M., 1998, ApJ, 504, 27
\bibitem[Maughan et al.(2005)]{maughan2005} Maughan, B. J., Jones, L. R., Ebeling, H., \& Scharf, C., 2005, MNRAS, 365, 509
\bibitem[McCarthy et al.(2003a)]{mccarthy2003a} McCarthy, I. G., Babul, A., Holder, G. P., \& Balogh, M. L., 2003a, 591, 515
\bibitem[McCarthy et al.(2003b)]{mccarthy2003b} McCarthy, I. G., Babul, A., Holder, G. P., \& Balogh, M. L., 2003b, 591, 526
\bibitem[Navarro, Frenk \& White(1997)]{nfw97} Navarro, J. F., Frenk, C. S. \& White, S. D. M., 1997, ApJ, 490, 493 
\bibitem[Novicki, Sornig, \& Henry(2002)]{novicki2002} Novicki, M. C., Sornig, M. A., \& Henry, J. P., 2002, ApJ, 123, 2413
\bibitem[Peebles(1993)]{peebles1993} Peebles, P. J. E., 1993, Principles of Physical Cosmology (Princeton Univ. Press)
\bibitem[Peebles, Daly, \& Juszkiewicz(1989)]{peebles1989} Peebles, P. J. E., Daly, R. A., \& Juszkiewicz, R., 1989, ApJ, 347, 563
\bibitem[Pierpaoli et al.(2003)]{pierpaoli2003} Pierpaoli, E., Borgani, S., Scott, D., \& White, M., 2003, MNRAS, 342, 163
\bibitem[Pointecouteau et al.(2004)]{pointecouteau2004} Pointecoutteau, E., Arnaud, M., Kaastra, J., \& de Plaa, J., 2004, A\&A, 423, 33
\bibitem[Ponman et al.(1999)]{ponman1999} Ponman, T. J., Cannon, D. B., \& Navarro, J. F., 1999, Nature, 397, 135
\bibitem[Pratt \& Arnaud(2005)]{pratt2005} Pratt, G. W. \& Arnaud, M., 2005, A\&A, 429, 791
\bibitem[Pratt, Arnaud, \& Pointecouteau(2006)]{pratt2006} Pratt, G. W., Arnaud, M., \& Pointecouteau, E., 2006, A\&A, 446, 429
\bibitem[Press \& Schechter(1974)]{press1974} Press,  W. H. \& Schechter,  P. I., 1974, ApJ, 187, 425
\bibitem[Raymond \& Smith(1977)]{raymondsmith77} Raymond, J. C. \& Smith, B. W., 1977, ApJS, 35, 419
\bibitem[Refregier, Valtchanov, \& Pierre(2002)]{refregier2002} Refregier, A., Valtchanov, I., \& Pierre, M., 2002, A\&A, 390, 1
\bibitem[Reiprich \& B\"{o}hringer(2002)] {reiprich2002} Reiprich, T. \& B\"{o}hringer, H, 2002, ApJ, 567, 716
\bibitem[Rosati et al.(2002)]{rosati2002} Rosati, P., Borgani, S., \& Norman, C., 2002, ARA\&A, 40, 539
\bibitem[Ruhl et al.(2004)]{ruhl2003} see Ruhl, J., et al., 2004, Proc. SPIE, 5498, 11
\bibitem[Sanderson et al.(2003)]{sanderson2003} Sanderson, A. J. R., Ponman, T. J., Finoguenov, A., Lloyd-Davies, E. J., \& Marketich, M., 2003, MNRAS, 340, 989
\bibitem[Seljak \& Zaldarringa(1996)]{seljak1996} Seljak, U. \& Zaldarringo, M., 1996, ApJ, 469, 437
\bibitem[Seljak(2002)]{seljak2002} Seljak, U., 2002, MNRAS, 337, 769
\bibitem[Seo \& Eisenstein(2003)]{seo2003} Seo, H.-J. \& Eisenstein, D. J., 2003, ApJ, 598, 720
\bibitem[Shuecker et al.(2003)]{shuecker2003} Shuecker, P., B\"{o}hringer, H., Collins, C. A., \& Guzzo, L., 2003, A\&A, 298, 867
\bibitem[Spergel et al.(2003)]{spergel2003} Spergel, D.N., Verde, L., Peiris, H.V. Komatsu, E., Nolta, M.R., Bennett, C.L., Halpern, M., Hinshaw, G., Jarosik, N., Kogut, A., Limon, M. Meyer, S.S., Page, L., Tucker, G.S., Welland, J.L., Wollack, E., \& Wright, E.L., 2003, ApJS, 148, 175
\bibitem[Spergel et al.(2005)]{spergel2005} Spergel, D. N., Bean, R., Dore', O., Nolta, M. R., Bennett, C. L., Hinshaw, G., Jarosik, N., Komatsu, E., Page, L., Peiris, H. V., Verde, L., Barnes, C., Halpern, M., Hill, R. S., Kogut, A., Limon, M., Meyer, S. S., Odegard, N., Tucker, G. S., Weiland, J. L., Wollack, E., \& Wright, E. L., ApJ, submitted, astro-ph/0603449
\bibitem[Sun et al.(2003)]{sun2003} Sun, M., Jones, C., Murray, S. S., Allen, S. W., Fabian, A. C., \& Edge, A. C., 2003, ApJ, 587, 619
\bibitem[Sunyaev \& Zeldovich(1972)]{sunyaev1972} Sunyaev, R. \& Zeldovich, Y., 1972, Comments Aprophys. Space Phys., 2, 66
\bibitem[Sunyaev \& Zeldovich(1980)]{sunyaev1980} Sunyaev, R. \& Zeldovich, Y., 1980, MNRAS, 190, 413
\bibitem[Tegmark, Taylor, \& Heavens(1997)]{tegmark1997} Tegmark, M., Taylor, A. N., \& Heavens, A. F., 1997, ApJ, 480, 22
\bibitem[Tyson(2002)]{lsst} Tyson, A. J., et al., 2002, Proc. SPIE Int. Soc. Opt. Eng. 4836, pp. 10--20, astro-ph/0302102 (see also www.lsst.org)
\bibitem[Verde, Haiman, \& Spergel(2002)]{verde2002} Verde, L., Haiman, Z., \& Spergel, D., 2002, ApJ, 581, 5
\bibitem[Viana \& Liddle(1996)]{viana1996} Viana, P. P. \& Liddle, A. R., 1996, MNRAS, 262, 1023
\bibitem[Vikhlinin et al.(1998)]{vikhlinin1998} Vikhlinin, A., McNamara, B. R., Forman, W., Jones, C., Quintana, H., \& Hornstrup, A., 1998, ApJ, 502, 558
\bibitem[Vikhlinin et al.(2004)]{vikhlinin2004} Vikhlinin, A., Markevitch, M., Murray, S. S., Jones, C., Forman, W., \& Van Speybroeck, L., 2004, ApJ, 628, 655
\bibitem[Voit et al.(2002)]{voit02} Voit, M. G, Bryan, G. L., Balogh, M.  L., \& Bower, R. G., 2002, ApJ, 576, 601
\bibitem[Wang et al.(2004)]{wang2004} Wang, S., Khoury, J., Haiman, Z., \& May, M., 2004, Phys. Rev. D., 70, 7013008 
\bibitem[Weller \& Battye(2003)]{weller2003} Well, J. \& Battye, R. A., 2003, New Astronomy Reviews, 47, 775
\bibitem[Weller, Battye, \& Kneissl(2002)]{weller2002} Weller, J., Battye, R. A., \& Kneissl, R., 2002, Phys. Rev. Lett., 88, 1301
\bibitem[White, Jones, \& Forman(1997)]{white1997}White, D. A., Jones, C., \& Forman, W., 1997, MNRAS, 292, 419
\bibitem[Willick \& Strauss(1998)]{willick1998} Willick, J. A. \& Strauss, M. A., 1998, ApJ, ApJ, 507, 64
\bibitem[Xu, Jin, \& Wu(2001)]{xu01} Xu, H., Jin, G \& Wu, X.-P. 2001, ApJ, 553, 78
\bibitem[Younger \& Bryan(2006)]{younger2006} Younger, J. D. \& Bryan, G. L., in preparation
\bibitem[Zaldarriaga, Spergel, \& Seljak(1997)]{zaldarriaga1997} Zaldarriaga, M., Spergel, D. N., \& Seljak, U., 1997, ApJ, 488, 1

\end{thebibliography}
\end{document}